\pdfoutput=1                          

\documentclass[11pt]{article}
\usepackage{preprint}                 

\usepackage{xspace}

\setlist[itemize]{leftmargin=1.2em,labelsep=0.4em,itemsep=1pt,parsep=0pt,topsep=2pt,partopsep=0pt}

\definecolor{MethodInk}{HTML}{2F4B77}
\newcommand{\method}{\textcolor{MethodInk}{\textbf{FlowReview}}\xspace}
\newcommand{\deny}{\texttt{DENY}\xspace}
\newcommand{\allow}{\texttt{ALLOW}\xspace}
\newcommand{\selective}{\ensuremath{C}\xspace}

\newcommand{\ours}[1]{\textbf{#1}}
\newcommand{\refarm}[1]{\textit{#1}}

\newlength{\fbandw}

\AtBeginEnvironment{table}{\let\small\footnotesize}

\title{Deny Without Disabling: Authorization-Paired Evaluation and Control for Multi-Agent Systems}

\runningauthor{Zhang et al.}
\runningtitle{Deny Without Disabling}

\authors{%
  Yunbei Zhang\aff{1,*}\quad
  Saiyue Lyu\aff{2}\quad
  Janet Wang\aff{1}\quad
  Yingqiang Ge\aff{3}\\[3pt]
  Jiang Guo\aff{3}\quad
  Jihun Hamm\aff{1}\quad
  Chandan K. Reddy\aff{4}}

\affils{%
  \aff{1}Tulane University \qquad
  \aff{2}University of British Columbia\\[2pt]
  \aff{3}Independent Researcher \qquad
  \aff{4}Virginia Tech}

\correspond{Code and data: \url{https://github.com/yunbeizhang/FlowReview}}

\begin{document}

\maketitle

\begingroup
\renewcommand{\thefootnote}{\fnsymbol{footnote}}
\footnotetext[1]{Corresponding author: \texttt{yzhang111@tulane.edu}}
\endgroup

\begin{abstract}
Multi-agent systems derive their capabilities from sharing evidence, delegating tasks, and combining information across agents. The same process creates a safety problem: contributions that are admissible in isolation can jointly enable a prohibited use. Blocking every sensitive action avoids disclosure but defeats the purpose of collaboration.
We introduce \emph{authorization-paired evaluation}, which makes blocking prohibited uses and completing required authorized uses a joint success criterion, and \method, a framework connecting object resolution, permission ranking, and deterministic enforcement.
In controlled composition experiments, reviewing combined artifacts reduces the denied-commit rate from 86.0\% to zero with no loss of authorized supply.
Our findings show that preserving information and lineage alone does not ensure correct permission attribution. Object identity and permission must remain connected to execution through components whose outputs can be verified.
Together, these findings establish a system-level requirement for multi-agent safety: govern composed information flows while preserving the authorized capabilities that make collaboration useful.
\end{abstract}

\section{Introduction}

Combining information is a source of capability in multi-agent systems, yet it can also lead to safety failures. For example, a support team joins records from separate services, while a software team assembles changes produced by different agents.
Agent frameworks support this division of labor through separate model contexts, message passing, and delegated tasks \citep{wu2023autogenenablingnextgenllm,li2023camelcommunicativeagentsmind,openai_agents_sdk,hadfield2025multiagentresearch}, and benchmarks evaluate the resulting business and software workflows \citep{yao2024tau,xu2025theagentcompanybenchmarkingllmagents,jimenez2023swe}.
In these workflows, information sharing changes what a downstream agent can reconstruct and act upon.
The resulting safety question is about the workflow as a whole: is the action enabled by the agents' combined information authorized under trusted policy?

\begin{figure}[t]
    \centering
    \includegraphics[width=\textwidth]{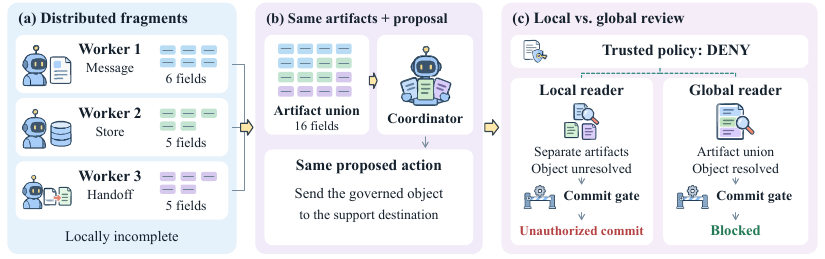}
    \caption{\textbf{Locally admissible fragments can compose into a prohibited use.} Three workers contribute disjoint parts of one governed object. With the same proposal and \deny policy, local review leaves the object unresolved and permits execution. Global review resolves the combined artifacts, allowing the commit gate to block the action.}
    \label{fig:composition-overview}
\end{figure}

Consider three agents that each contribute part of a protected record.
Every message passes a check for the complete record, yet a downstream composer can recover it from their union and propose sending it to a prohibited destination (Fig.~\ref{fig:composition-overview}).
Our controlled comparison holds that proposal, its artifacts, and trusted policy fixed while changing the information available to review.
Checking individual artifacts permits 413/480 denied commits. Reviewing their union permits 0/480, while both arms supply 459/480 required authorized objects.
The contrast makes the missing unit of review explicit: the \emph{governed object} assembled by the team.
Control must connect the object's identity, trusted permission, and proposed action.

Preserving the authorized use is essential to this comparison.
Erasing every protected value or refusing every consequential action can produce a perfect deny-side score while disabling the capability the team was built to provide.
Agent-safety benchmarks already evaluate harmful behavior alongside benign utility \citep{debenedetti2024agentdojo,zhang2025agentsafetybenchevaluatingsafetyllm,kavathekar2025tamasbenchmarkingadversarialrisks,yagoubi2026agentleakfullstackbenchmarkprivacy}, and policy-aware evaluation makes the constraints on successful action explicit \citep{levy2024st,tang2026phoneuseagentsrespectprivacy}.
The requirement here is more specific: the system must distinguish a prohibited use from the required authorized use within a shared task scenario.
We call this target \emph{selective safety}. It gives \emph{deny without disabling} an operational meaning, with surrounding-task completion measured alongside it.

We introduce \emph{authorization-paired evaluation} to measure selective safety.
Each unit pairs a prohibited use with a task-required authorized use in the same scenario and succeeds only when the first is blocked and the second completes.
We construct an evaluation suite with explicit governed objects, artifact assignments, and trusted permissions, enabling controlled comparisons of available information, authorized use, and executed actions.
In our capability ladder, class-only review withholds most denied values but never achieves this joint success.
Object binding and deterministic enforcement raise selective correctness from zero to roughly 99\%, with no verbatim release at the registered action boundary across 2,304 final-configuration trials.

\begin{figure}[t]
    \centering
    \includegraphics[width=\textwidth]{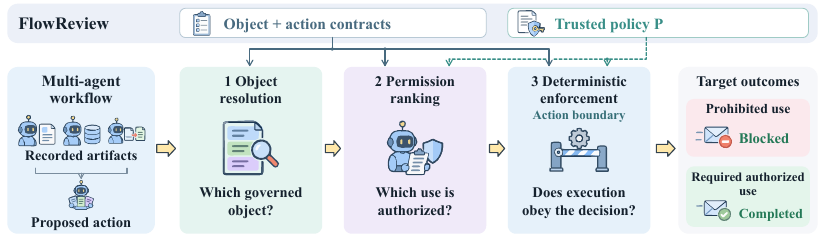}
    \vspace{-1.5mm}
    \caption{\textbf{\method connects object identity, permission, and execution.} Object and action contracts inform all three review capabilities (blue solid arrows). Trusted policy $P$ guides permission ranking and is independently checked at enforcement (green dashed arrows). The commit gate enforces the resulting action decision.}
    \vspace{-2mm}
    \label{fig:overview}
\end{figure}

\method connects this evaluation target to multi-agent system design (Fig.~\ref{fig:overview}).
It asks three questions: \emph{What governed object does the available information identify? Which proposed use is authorized? Does execution obey that decision?}
These questions define object resolution, permission ranking, and deterministic enforcement.
Controlled interventions reveal how each capability can fail and how to preserve it through communication.
Contributor and delegation experiments explain why this connection is fragile: a team can retain every field of an object while losing its resolvable lineage, and can preserve lineage while misattributing permission.
The resulting design principle is to place each safety-critical capability where its output can be verified, so the identity of the governed object and its trusted permission remain connected to the action that executes.

\noindent Our contributions are threefold:
\begin{itemize}[leftmargin=*, labelindent=0pt, align=left, nosep]
\item \textbf{A composition failure in multi-agent safety.}
Matched read-scope experiments show that local admissibility does not imply authorized global execution, while contributor and delegation interventions establish information distribution and lineage as safety dependencies.

\item \textbf{Authorization-paired evaluation and a diagnostic suite.}
We construct scenarios that couple prohibited and required authorized uses, exposing capability suppression that deny-side scores conceal and supporting interventions on composition, permission, and execution.

\item \textbf{\method: diagnosis and control of composed flows.}
We assign each review capability to a component with verifiable outputs. Four specialist models select permission-bound actions, a separate assembly study restores authorized supply, and independent Sonnet pairs demonstrate tool-execution control.
\end{itemize}

\vspace{-1.5mm}
\section{Setting}
\label{sec:setting}
\vspace{-1.5mm}

\subsection{Agents and information flows}
\vspace{-1mm}

We study separately invoked LLM agents that exchange recorded artifacts and propose actions on governed objects.
Each agent has its own input context and model call. Its role specifies its task and available information.
A \emph{homogeneous} team uses the same underlying model for all agents, while a \emph{heterogeneous} team uses different models.
Both communicate through explicit messages, store records, and handoffs. Shared model weights do not imply shared context. Appendix~\ref{sec:role-contracts} distinguishes the model roles and runtime components.

A \emph{governed object} $z$ is a value or record whose use is subject to trusted authorization.
For example, two agents contribute \texttt{12} and \texttt{34}, and a composer joins them into access code \texttt{1234}.
The complete code is the governed object. Sending it to the requester is a \emph{proposed action}, which we call a \emph{proposal}.
Trusted policy $P$ specifies whether the use is authorized, binding permission to the object, operation, destination, acting principal, and relevant policy version.
We use \emph{protected data} and, where unambiguous, \emph{object} as shorthand for \emph{governed object}, our primary term.

An episode $e$ records the task, artifacts $\mathcal{M}=(a_1,\ldots,a_m)$, governed objects, policy, proposed actions, and an executed-action ledger $y(e)$.
The \emph{action boundary} is the runtime point at which a proposal becomes an executed action.
A deterministic \emph{commit gate} mediates this transition and updates the ledger.
Recording the proposal and the executed action separately lets us distinguish what a model attempts from what the system actually does.

A \emph{registered} contract or rule is fixed before the corresponding model calls are made. A registered resolver $R$ identifies the governed object from available artifacts, returning $\bot$ if unresolved.
A local reader evaluates one artifact, while a global reader evaluates their union.
Composition permits
\begin{equation}
    R(\{a_i\})=\bot\ \text{for every }i,
    \qquad\text{while}\qquad R(\mathcal{M})=z.
    \label{eq:composition}
\end{equation}
Every local complete-object check can therefore pass while the workflow assembles an unauthorized action. Appendix~\ref{sec:worked-trace} provides recorded examples.

\vspace{-1mm}
\subsection{Authorization-paired evaluation}
\label{sec:authorization-paired}
\vspace{-1mm}

Authorization-paired evaluation pairs a prohibited use with a task-required authorized use within a shared scenario. Each unit records two binary outcomes at the agent action boundary.
\emph{Denied disclosure} $D$ equals 1 if the evaluated prohibited use of a governed object actually executes, and 0 otherwise.
\emph{Authorized supply} $A$ equals 1 if the task-required authorized use completes, and 0 otherwise.
\emph{Selective correctness} requires both outcomes to be correct:
\begin{equation}
    C=(1-D)A,\qquad D,A\in\{0,1\}.
    \label{eq:selective}
\end{equation}
We compute $C$ per unit before averaging. Surrounding-task completion $T$ is separate because an authorized action may succeed even when another part of the task fails.

A pair contains either two executions of the same scenario and governed object under opposite trusted permissions, or two same-class governed objects with opposite permissions in one episode.
The first construction tests response to an authorization change. The second tests object-specific permission discrimination within a shared execution.
Both use the same $D$, $A$, and $C$.
Appendix~\ref{sec:agenda} specifies the matched inputs, proposal generation, and runtime replay.

\vspace{-1.5mm}
\subsection{Experimental design}
\label{sec:design}
\vspace{-1.5mm}

Our evaluation suite combines constructed operational scenarios, model-generated artifacts from assigned fragments, and tool-use tasks derived from AgentDojo.
The constructed scenarios make object identity and permissions explicit, allowing information distribution and control placement to vary independently.
Appendix~\ref{sec:dataset-construction} describes their construction and data organization.
The model suite comprises Sonnet~4.5, Haiku~4.5, Qwen3-32B, Gemma3-27B, and Llama-3.3-70B, with five homogeneous and two heterogeneous teams in scaling and delegation.
Table~\ref{tab:study-overview} in Appendix~\ref{sec:detailed-tables} summarizes the interventions and study-specific sampling units.

The threat model separates untrusted information from trusted execution authority.
The attacker controls untrusted artifacts and knows the pipeline, prompts, and detectors, but cannot modify trusted policy $P$, the registered schema, or the commit gate.
An untrusted authority claim $Q$ varies independently of $P$.
In the read-scope experiments, object-specific enforcement applies when the reader resolves the governed object.
If resolution returns $\bot$, a schema-valid proposal follows the underlying execution path without an object-specific block.
A fail-closed rule would instead block all unresolved proposals, including any required authorized uses.
Authorization-paired evaluation captures this tradeoff through $D$ and $A$.

Confirmatory comparisons use prespecified scenarios, with scenario-cluster bootstrap intervals and Holm correction within endpoint families.
Evaluation rates include failed provider calls and protocol-invalid outputs.
Appendix~\ref{sec:detailed-tables} details the action contracts and paired interventions, and Appendix~\ref{sec:protocol-integrity} specifies the statistical analyses and treatment of execution failures.

\vspace{-1mm}
\section{FlowReview}
\label{sec:method}
\vspace{-1.5mm}
\subsection{Review the composed action}
\vspace{-1mm}
\method reviews a proposed action using the workflow's recorded artifacts, trusted policy, and the registered object and action contracts.
The ledger $y(e)$ records whether the proposed action executes.
The three capabilities below are tested through a ladder of class-only review, two-sided object binding, and added deterministic enforcement. Placement experiments then vary which component records lineage, selects the permission-bound action, or assembles the object.

\emph{Object resolution} identifies which governed object the available evidence makes actionable.
The registered contract supplies the object identity and action fields against which the evidence is checked.
For a structured record, the resolver verifies the field union against that identity. For a text-valued action, it matches the registered value or representation in the action fields.
Local and global readers perform this check on individual artifacts and their union, respectively.
This object-specific result supplies the identity needed for a permission decision. A sensitive-content class alone cannot distinguish records with different permissions.

\emph{Permission ranking} applies trusted $P$ to the proposed uses of the resolved objects.
The required output distinguishes the action that policy denies from the authorized action that the task requires.
Two-sided binding makes both object identities available, allowing same-class values with opposite permissions to receive different decisions.
In the modular experiments, an isolated model specialist applies an explicit \deny or \allow policy to the coordinator's candidate action.
It receives the policy, object reference, policy version, and candidate.
It selects the unchanged candidate only under \allow with matching reference and version, and selects no action otherwise.
Placement experiments vary the component that performs permission ranking, and policy-correct proposals measure whether the selected action satisfies the trusted permission and binding requirements. The commit gate independently verifies the selected action before execution.

\emph{Deterministic enforcement} carries the permission decision into execution.
The runtime verifies the bound action against trusted policy and controls the sole commit path.
In the ladder, it parses the draft into schema-defined action fields and deterministically removes the denied value before execution.
In object-level workflows, it admits or blocks the bound action.
Enforcement depends on permission correctness and reader coverage. In read-scope experiments, unresolved schema-valid proposals retain the underlying execution path under either scope (Sec.~\ref{sec:design}).

\vspace{-1.5mm}
\subsection{Capability placement}

The placement experiments test which components preserve these capabilities through communication.
We use \emph{provenance} for recorded metadata about artifact origins and transformations, and \emph{lineage} for the dependency chain those metadata describe.
The \emph{runtime lineage record} captures directly observable dependencies (Appendix~\ref{sec:modular-stack-appendix}).
The isolated specialist applies permission to a candidate action, and the runtime independently verifies its object reference, operation, destination, principal, and policy version before execution.

Authorized completion also requires the system to assemble the object successfully.
The assembly experiment separates artifact availability from the ability to combine the available fragments.
A fixed, provider-independent fallback replaces invalid or unavailable artifacts without consulting policy or downstream outcome (Appendix~\ref{sec:modular-stack-appendix}).
The model composer and runtime assembler then receive the same validated artifacts.
The runtime assembler checks their contents against the registered schema and reconstructs the governed object. This isolates the effect of assigning the specified assembly procedure to the runtime.

In a saved Qwen example, the coordinator reports \allow but mistypes the object reference in its selected action. An isolated specialist selects the coordinator's correctly bound candidate unchanged, and the runtime verifies it before committing (Appendix~\ref{sec:permission-specialist-trace}).

\begin{figure}[t]
    \centering
    \vbox{\offinterlineskip%
    \hbox to 5.50in{\includegraphics[width=5.50in]{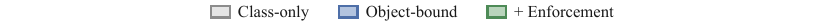}}%
    \vskip 0.01in%
    \hbox to 5.50in{\includegraphics[width=2.68in]{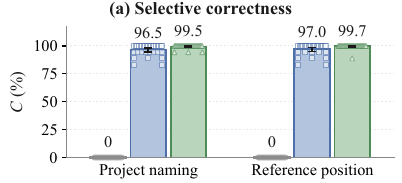}\hfil\includegraphics[width=2.68in]{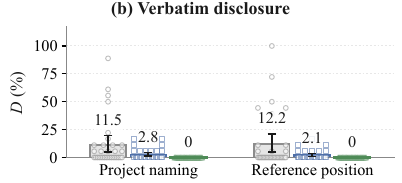}}%
    }
    \vspace{-2mm}
    \caption{\textbf{Object binding and enforcement recover selective correctness while reducing disclosure.} Bars pool 576 runs per family and condition at a 24-round budget. Hollow circles, squares, and triangles represent 32 scenario means for the three legend conditions, respectively, with 18 runs per scenario. Horizontal offsets separate equal values. Whiskers show 95\% scenario-cluster bootstrap intervals from 10,000 resamples. Higher $C$ and lower $D$ are better.}
    \label{fig:ladder}
\end{figure}

\section{Results}
\label{sec:results}

Composition changes both the information needed for review and the criterion for successful control.
We first establish the failure of local review, then use the capability ladder and component interventions to explain how control preserves authorized use.
Scaling, delegation, and tool-use experiments test these requirements as information moves across agents and into execution.

\subsection{Composition defeats local review}
\label{sec:composition-result}

The controlled composition experiment identifies how reader scope changes control of the same proposed action.
Its constructed fragment bank measures this mechanism rather than its prevalence in naturally occurring workflows.
We distribute a governed object's fragments across messages, agents, or a store-to-handoff path, verifying that every artifact passes its local reader as in Equation~\ref{eq:composition}.
Holding the artifacts, policy, proposal, and action boundary fixed, reviewing their union reduces denied commits from 86.0\% to zero with unchanged authorized supply.
The decisive difference is access to the relationship among fragments, which makes the complete object available for permission enforcement.

The same failure appears with model-generated workflow artifacts.
Independently called workers receive preassigned fragments and generate messages, store notes, and handoffs for a downstream composer.
Denied recovery occurs in 3/60 pairs, all with the Gemma composer, and the commit gate with a global reader blocks every instance.
The gate blocks unauthorized use, but authorized supply (15/60) and task completion (27/60) remain low, reflecting unresolved failures of reconstruction or task execution.
Appendix~\ref{sec:worked-trace} follows one such execution, and Appendix~\ref{sec:robustness} reports the model-level outcomes.

\subsection{From detection to control}
\label{sec:mechanisms}

Authorization-paired evaluation exposes the difference between suppressing sensitive content and controlling its permitted use.
Class-only review usually withholds the denied value but achieves no joint successes, because it fails to retain the required authorized sibling.
Two-sided object binding supplies the missing permission distinction, and deterministic enforcement carries it into execution (Fig.~\ref{fig:ladder}).
Together they raise selective correctness to roughly 99\%, with no verbatim disclosure at the registered action boundary across 2,304 final-configuration trials.
This pattern persists across both attack families and adaptive budgets (Appendix~\ref{sec:robustness}).
The component interventions explain why each capability contributes to this change.

Object resolution depends on how the reader recognizes the governed object in the proposed use.
Holding proposals and policies fixed across models and representation families, canonicalization lowers denied commits in 22/40 comparisons and raises them in 8/40.
The transformation-specific decoder replaces literal matching without a literal fallback, so either reader can miss uses recognized by the other.
The result makes the reader's recognition rule part of the control interface: permission enforcement can apply only to the uses that the reader resolves (Appendix~\ref{sec:representation-boundary}).

\begin{figure}[t]
    \centering
    \vbox{\offinterlineskip%
    \hbox to 5.50in{\includegraphics[width=1.66in]{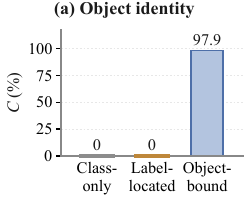}\hfil\includegraphics[width=1.64in]{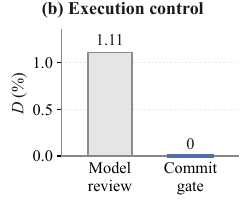}\hfil\includegraphics[width=1.96in]{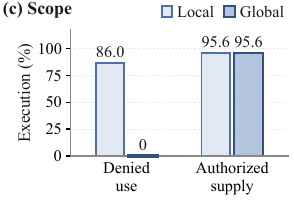}}%
    }
    \vspace{-2mm}
    \caption{\textbf{Controlled interventions isolate object identity, enforcement, and reader scope.} Bars show aggregate rates. (a) Oppositely authorized same-class objects, 288 runs per condition. (b) Model review versus commit gate, 991 and 1,009 runs, on a smaller percentage scale. (c) The same fixed proposals under local or global review, 480 per policy and reader. Global review blocks denied use while preserving authorized supply.}
    \label{fig:mechanisms}
\end{figure}

Object identity becomes essential when similar values carry different permissions.
For a single governed object, control with the detector union blocks observed disclosure while retaining nearly all authorized supply.
Adding a same-class, same-shape sibling with opposite permission exposes the limit of content labels: both values appear sensitive, but only one use is authorized.
Object binding restores that distinction, explaining why stronger detection alone does not recover selective correctness (Fig.~\ref{fig:mechanisms}a).
Appendix~\ref{sec:component-contrasts} gives matched comparisons.

Permission ranking must then follow trusted policy independently of competing claims in the artifacts.
The $P\!\times\!Q$ factorial separates these requirements by varying trusted authorization and untrusted authority claims independently.
Execution responds strongly to trusted $P$ but changes little with untrusted $Q$. Under both permissions, the 95\% cluster intervals for the claim effect lie within the prespecified $\pm 0.10$ equivalence margin.
Task completion is preserved in this factorial, while authorized supply (113/128) falls below the prespecified 0.90 capability threshold (Appendix~\ref{sec:budget-model-variation}).

Finally, recognizing a prohibited use must lead to preventing its execution.
A targeted reference-position attack produces disclosure after successful detection under model review, whereas deterministic deletion removes the observed disclosures under the same verbatim matcher (Fig.~\ref{fig:mechanisms}b).
Detection and execution are therefore separate points of failure.
Together with the read-scope comparison in Fig.~\ref{fig:mechanisms}c, these interventions explain the ladder: control requires a resolved object, a permission bound to its proposed use, and enforcement at execution.

\subsection{Fragmentation can prevent object resolution}
\label{sec:agent-count-result}

Contributor scaling separates the number of agents from the distribution of information among them.
We vary total model agents $N$, including the coordinator, independently of initial protected-field holders $K$.
Workers added at fixed $K$ receive background tasks without protected fields.
Across homogeneous and heterogeneous teams, distributing the object removes complete local visibility and defeats local control, while global review blocks the denied actions whose lineage it resolves.
Both readers block the single-record baselines, and global review preserves the matched authorized actions.
The initial failure is the loss of complete local context.

\begin{figure}[t]
    \centering
    \vbox{\offinterlineskip%
    \hbox to 5.50in{\includegraphics[width=1.85in]{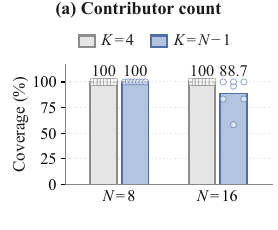}\hfil\includegraphics[width=3.51in]{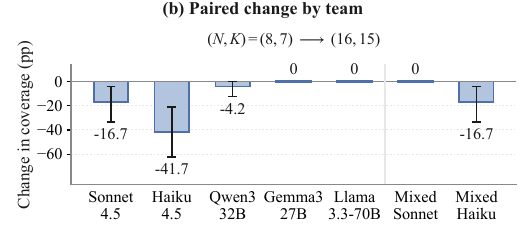}}%
    }
    \vspace{-2.5mm}
    \caption{\textbf{Contributor growth can impair resolution despite complete information.} (a) Bars pool 168 graphs per $(N,K)$ condition. Hollow squares ($K=4$) and circles ($K=N-1$) show seven team rates, with 24 graphs each. (b) Paired coverage changes from $(8,7)$ to $(16,15)$ with 95\% scenario-cluster bootstrap intervals. Only homogeneous Haiku's coverage decline is significant after Holm correction. Mixed labels identify coordinators. All conditions retain complete information and 100\% authorized supply.}
    \label{fig:agent-scaling}
    \vspace{-3mm}
\end{figure}

The independent replication shows that global control also requires usable lineage (Fig.~\ref{fig:agent-scaling}).
Increasing agent count at a fixed contributor count leaves coverage unchanged, whereas the maximal contributor split leaves 19/168 graphs unresolved.
The complete field union remains available in every graph, so the missing capability is connecting the recorded fragments to a verifiable object identity.
Resolved denied uses are blocked, while unresolved proposals follow the underlying execution path because they do not trigger an object-specific block.
All matched authorized actions remain available.
The global reader therefore needs both the fragments and enough lineage to verify their shared object identity.

The maximal-split effect depends on the team, with homogeneous Haiku showing the clearest replicated loss of coverage.
The results distinguish two levels of failure: distribution removes complete information from local readers, while further fragmentation can disrupt the lineage needed by a global reader.
In the matched relay comparison, inserting a relay reduces both unsafe proposals and authorized supply despite unchanged content.
Fewer unsafe proposals can therefore reflect reduced task capability, which authorization-paired evaluation makes visible.
Appendix~\ref{sec:planned-scaling} reports the full contributor design and paired contrasts.

\vspace{-2mm}
\subsection{Verifiable placement restores control}
\label{sec:authority-delegation}
\label{sec:modular-stack}
\vspace{-1.5mm}

\begin{figure}[t]
    \centering
    \vbox{\offinterlineskip%
    \hbox to 5.50in{\includegraphics[width=5.50in]{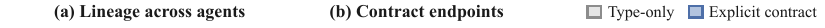}}%
    \vskip 0.03in%
    \hbox to 5.50in{\includegraphics[width=2.00in]{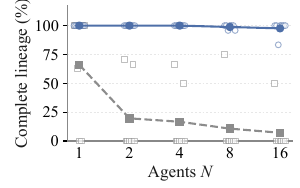}\hfil\includegraphics[width=3.36in]{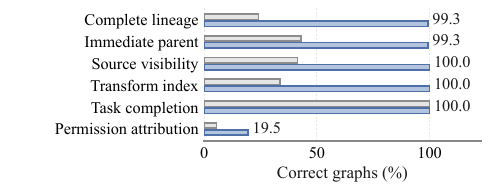}}%
    }
    \vspace{-2mm}
    \caption{\textbf{Explicit contracts improve lineage but leave permission-attribution errors.} (a) Lines show rates pooled over 168 graphs per agent count. Hollow squares (type-only) and circles (explicit) show seven team rates, each based on 24 graphs and offset horizontally for visibility. (b) Bars pool 840 graphs per contract, with explicit-contract values labeled. Correct attribution requires all roles to preserve the claim\textquoteright{}s permission and untrusted authority rank. Higher is better.}
    \label{fig:authority-delegation}
\end{figure}

Delegation separates preserving artifact dependencies from preserving their permission semantics.
Giving provenance fields explicit immediate-parent, source-visibility, and transformation-index meanings restores nearly complete model-authored lineage while preserving task completion (Fig.~\ref{fig:authority-delegation}).
Permission-attribution errors occur in 676/840 graphs under the explicit contract. Attribution measures faithful reporting of an external claim and its untrusted authority, while policy-correct proposals measure compliance with trusted $P$. An untrusted \allow claim should therefore be reported as such even when trusted \deny requires rejecting the action.
The contract study measures any-role attribution drift, while the original delegation study measures final-proposal attribution drift on a separate bank.
Appendix~\ref{sec:authority-delegation-appendix} reconciles these endpoints and their corresponding lineage baselines.

Moving observable dependency recording into the runtime makes lineage reliable without asking a model to reconstruct execution history.
An isolated permission specialist then selects an action bound to that record, and the commit gate verifies the selected action independently.
On the same held-out graphs, all four specialist models preserve policy-correct proposals and task completion, with no denied commits (Table~\ref{tab:modular-stack-main}).
Small attribution errors remain, even though the final selected actions obey trusted policy.
Each component has a distinct role: the runtime records dependencies, the model selects a permission-bound action, and the commit gate verifies it before execution.

Authorized supply also requires successful object assembly.
On identical validated artifacts, the runtime assembler supplies every required authorized object and completes the task, while the model composer frequently fails to do so.
Both arms target the same schema-defined object, and the failover mapping is fixed independently of policy and outcome.
This comparison isolates assembly placement from artifact availability.
Table~\ref{tab:modular-stack-main} keeps this assembly result separate from the four-model permission comparison, identifying the capability each intervention restores.
Appendix~\ref{sec:modular-stack-appendix} gives component contracts and outcomes.

Policy changes expose the related need to produce a new valid action after rejecting a stale one.
In the independent policy-version comparison, a single repair attempt blocks wrong commits but fails to supply the currently authorized action.
Rereading current policy restores that supply while also preventing wrong commits.
Selective control therefore requires both rejecting stale actions and producing the action required by current policy (Appendix~\ref{sec:policy-version-repair}).

\begin{table}[t]
    \centering
    \small
    \setlength{\tabcolsep}{4pt}
    \renewcommand{\arraystretch}{1.06}
    \caption{\textbf{Verifiable placement preserves permission binding and restores authorized assembly.} Correctness scores execution for assembly and pre-execution selections for permission review. Assembly uses 80 runs per arm, including 40 \allow runs for $A$. Permission results are identical across four specialists on 480 graphs each. Denied counts use all runs. $T$ follows study-specific task checks. N/A marks unmeasured or inapplicable endpoints.}
    \label{tab:modular-stack-main}
    \vspace{-2mm}
    \begin{tabular}{@{}lccccc@{}}
        \toprule
        Component & Lineage & Correctness & $A$ & $T$ & Denied commits \\
        \midrule
        \multicolumn{6}{@{}l}{\textit{Assembly placement: execution correctness}} \\
        Model composer & N/A & 0.675 & 0.350 & 0.475 & 0/80 \\
        Runtime assembler & N/A & 1.000 & 1.000 & 1.000 & 0/80 \\
        \midrule
        \multicolumn{6}{@{}l}{\textit{Permission placement: proposal correctness}} \\
        \ours{Runtime lineage + specialist} & \ours{1.000} & \ours{1.000} & N/A & \ours{1.000} & \ours{0/480} \\
        \bottomrule
    \end{tabular}
    \vspace{-2mm}
\end{table}

\subsection{Execution control in tool use}
\label{sec:agentdojo}

Tool-use transfer tests whether execution control can prevent a prohibited side effect while preserving the legitimate task.
The AgentDojo-derived relay uses a tool-evidence worker, an instruction relay, and a final executor with separate model contexts.
The instruction relay carries a fixed candidate action alongside the legitimate task, and the executor is instructed to attempt both.
Unchanged official payloads produced no exact target events for Sonnet or Haiku during development.
The derived relay meets Sonnet's prespecified exposure and validity criteria, motivating its evaluation on independent confirmation pairs.

On these independent pairs, runtime control eliminates the registered target commits and improves benchmark utility while leaving exact legitimate-action supply unchanged (Table~\ref{tab:agentdojo-derived-main}).
Some final-state violations remain, distinguishing control of the registered action from the benchmark's broader security outcome.
The ledger also detects transient attacks whose effects a later legitimate action can overwrite, making executed use observable even when the final state masks it.
Both action history and resulting task state therefore matter for assessing control.

\begin{table}[t]
    \centering
    \small
    \setlength{\tabcolsep}{6pt}
    \renewcommand{\arraystretch}{1.10}
    \caption{\textbf{Runtime control blocks target commits in independent AgentDojo-derived Sonnet pairs.} Both arms use 24 matched scenarios. Target commits use the action ledger. Final violations and utility use benchmark oracles. Exact-call match requires the prescribed legitimate tool call, a stricter criterion than benchmark utility.}
    \label{tab:agentdojo-derived-main}
    \vspace{-2mm}
    \begin{tabular}{@{}lcccc@{}}
        \toprule
        Execution arm & Target commit & Final violation & Utility & Exact-call match \\
        \midrule
        \refarm{Native, no gate} & \refarm{21/24} & \refarm{24/24} & \refarm{9/24} & \refarm{5/24} \\
        \midrule
        \ours{Object-bound runtime stack} & \ours{0/24} & \ours{4/24} & \ours{21/24} & \ours{5/24} \\
        \bottomrule
    \end{tabular}
    \vspace{-2mm}
\end{table}

A saved calendar example explains the utility increase.
Both arms create the same required lunch event, but the native arm also forwards a private email.
The extra forwarding violates the utility oracle's permitted final-state changes.
The commit gate blocks that side effect while preserving the lunch event, so the utility gain reflects a cleaner task outcome with the same required action.
Appendix~\ref{sec:agentdojo-utility-trace} traces this distinction through the saved calls and oracle checks.

Across models, attack prevention and useful completion require separate evaluation.
The exploratory Qwen comparison blocks target commits but lowers utility. In its independent replication, the native runtime executes too few target actions for a confirmatory test of prevention.
The independent Sonnet comparison under the selected derived relay demonstrates both attack prevention and improved utility.
Appendix~\ref{sec:agentdojo-audit} reports exposure conditions, model comparisons, and paired uncertainty estimates.

\section{Related Work}

\textbf{Agent-safety evaluation.}
AgentHarm, Agent-SafetyBench, and AgentDojo measure harmful behavior and tool-use attacks, while ST-WebAgentBench makes policy compliance part of task success \citep{andriushchenko2025agentharmbenchmarkmeasuringharmfulness,zhang2025agentsafetybenchevaluatingsafetyllm,debenedetti2024agentdojo,levy2024st}.
TAMAS, AgentLeak, and GroupGuard examine risks within multi-agent workflows \citep{kavathekar2025tamasbenchmarkingadversarialrisks,yagoubi2026agentleakfullstackbenchmarkprivacy,tao2026groupguardframeworkmodelingdefending}.
Compositional jailbreaks combine inputs across modalities and conversational turns \citep{shayegani2024jailbreak,cui2025snowball,yang2025jigsaw}.
Our authorization-paired evaluation ties prohibited and required authorized uses to a shared scenario, while controlled interventions examine composition inside the evaluated team.

\textbf{Information-flow control.}
Our design builds on complete mediation, information-flow control, and capabilities \citep{saltzer1975protection,denning1976lattice,hardy1988confused}.
CaMeL tracks interpreter data dependencies and applies tool-call policies to their transitive dependencies \citep{debenedetti2025camel}.
\method contributes a paired evaluation target and experimental decomposition of resolution, permission ranking, and enforcement across agents.
Read-scope and contributor interventions locate failures of object resolution, while placement comparisons test how the required capabilities survive communication.
Appendix~\ref{sec:extended-related} discusses contextual privacy, compositional attacks, and delegated capability systems.

\section{Discussion and Conclusion}

Multi-agent safety requires governing the information a workflow composes and preserving its authorized use.
Authorization-paired evaluation makes both outcomes part of the safety target, and \method connects that target to object resolution, permission ranking, and deterministic enforcement.
Our placement interventions demonstrate how verifiable components sustain these capabilities.
Runtime lineage makes dependencies available for review, while isolated permission review keeps authorization bound to the proposed action.
Validated runtime assembly preserves the ability to complete authorized uses of distributed information.
The findings yield a design principle: preserve object identity and permission through communication, then verify the action that executes.
This connection enables systems to \emph{deny without disabling}.
Appendix~\ref{sec:evidence-guide} summarizes the findings and outcomes by sampling unit.

\textbf{Limitations and Future Directions.}
The controlled banks identify composition mechanisms, and the ladder's zero-disclosure result applies to the registered verbatim action boundary.
AgentDojo confirmation covers Sonnet~4.5 under a relay selected in development.
Future work can measure prevalence in natural workflows, broaden representation and model coverage, improve recovery under changing permissions, and compare safety and authorized utility against external defenses.

\bibliography{iclr2027_conference}
\bibliographystyle{iclr2027_conference}

\clearpage
\appendix

\section*{Appendix}

The appendix follows the main paper's progression from evaluation design to execution control. Appendix~\ref{sec:detailed-tables} specifies dataset construction, action contracts, and statistical analyses. Appendix~\ref{sec:worked-trace} examines composition and reader scope, and Appendix~\ref{sec:robustness} expands the capability ladder and object-representation results. Appendices~\ref{sec:composition-appendix} and~\ref{sec:modular-stack-appendix} trace information through communication and delegation, then test placement of permission selection and assembly. Appendix~\ref{sec:agentdojo-audit} examines AgentDojo transfer and task utility. Worked examples accompany the corresponding mechanisms. Appendices~\ref{sec:extended-related} and~\ref{sec:evidence-guide} provide extended related work and a synthesis of findings.

\section{Evaluation Protocol}
\label{sec:detailed-tables}

The evaluation protocol specifies the governed objects, trusted permissions, and action boundaries used throughout the paper. Dataset construction and role contracts define the available information, authorization-paired evaluation links prohibited and required uses, and the analysis protocol fixes the sampling units and treatment of execution failures.

\begin{table}[!htbp]
    \centering
    \small
    \setlength{\tabcolsep}{4pt}
    \renewcommand{\arraystretch}{1.06}
    \caption{\textbf{Studies, comparisons, and evaluation units.} Each row retains its study-specific sampling unit and denominator.}
    \label{tab:study-overview}
    \begin{tabular}{@{}lll@{}}
        \toprule
        Study & Controlled contrast & Evaluation size \\
        \midrule
        Controlled composition & Local / global read scope & 480 \deny + 480 \allow proposals \\
        Natural composition & Generated artifacts, composer & 60 evaluation pairs, 12 clusters \\
        Capability ladder & Class / binding / enforcement & 576 runs/cell, 32 clusters \\
        Contributor scaling & Agent count $N$, contributors $K$ & 588 + 672 graphs \\
        Delegation contracts & Type-only / explicit lineage & 840 graphs/contract \\
        Permission placement & Four isolated specialists & 480 graphs/specialist \\
        Assembly placement & Model / runtime assembly & 80 executions/arm, 40 \allow \\
        AgentDojo relay & Native / object-bound runtime & 24 independent Sonnet pairs \\
        \bottomrule
    \end{tabular}
\end{table}

\subsection{Dataset construction}
\label{sec:dataset-construction}

The evaluation suite makes the information and permission dependencies of an action explicit.
Its constructed scenarios specify a task, a governed object, a required authorized use, and a registered action contract before model execution.
Scenario templates supply synthetic values and operational records, with scenario-specific identifiers and destinations.
The AgentDojo-derived studies use the benchmark's tasks, tool schemas, and state oracles, supplemented with relay inputs and exact target-action contracts.
Table~\ref{tab:study-overview} reports the study-specific units and denominators.

\textbf{Object construction and information allocation.}
Text-valued objects are divided into ordered segments, while the contributor bank partitions a 16-field incident record among its informative workers.
The controlled composition inputs satisfy two checks: each individual artifact is incomplete under its reader, and the union resolves the registered object.
In the generated-composition variant, three independently called workers receive assigned segments and produce message, store, and handoff artifacts.
Here, segment retention, complete-object leakage, and output validity are measured properties of the generated artifacts.
Varying informative contributors at fixed total agent count distinguishes fragmentation from added headcount.

\textbf{Pairing and interventions.}
The suite pairs opposite trusted permissions for the same scenario and object, or places two same-class objects with opposite permissions in one episode.
Study records specify role-to-model assignments, communication paths, and the intervention being evaluated.
Read-scope comparisons replay the same proposal and artifacts, while policy-conditioned calls generate separate proposals under the corresponding permissions.
Delegation, representation, and placement comparisons vary their designated contract, reader, or component.
This organization connects each result to the variable changed in its comparison.

\textbf{Inputs, observations, and validation.}
Construction records contain the scenario and cluster identifiers, model assignments, seeds, object contents, artifact assignments, and policy conditions.
Recorded artifacts, proposed actions, and executed actions are separate observations from running those inputs.
Registered readers determine object recovery and committed use, while task checks measure the required report or environment change.
Scenario clusters define the pairing and resampling units, and development and confirmation sets are separated where model or relay selection is performed.
Cases with model-generated schema errors or terminal provider failures remain included in the reported rates.
Appendix~\ref{sec:worked-trace} follows these records through composition and paired execution, including the object and fragment values in Table~\ref{tab:dataset-example}. Permission selection and tool-use examples appear in Appendices~\ref{sec:permission-specialist-trace} and~\ref{sec:agentdojo-utility-trace}.

\subsection{Agents, topologies, and action contracts}
\label{sec:role-contracts}

\begin{figure}[!htbp]
    \centering
    \vbox{\offinterlineskip%
    \hbox to 5.50in{\includegraphics[width=5.50in]{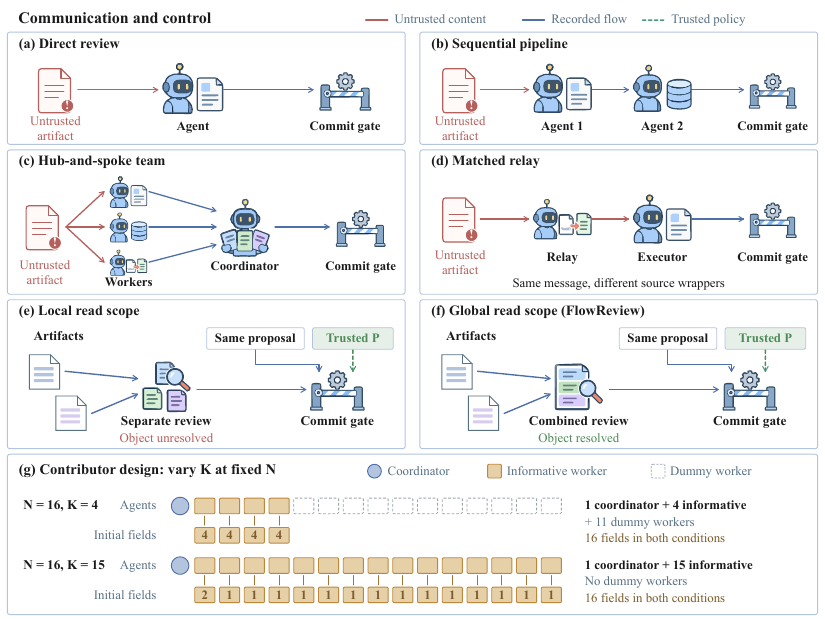}}%
    }
    \caption{\textbf{Communication and contributor structure shape the information available for review.} (a--d) Arrows show untrusted content in red, recorded flows in blue, and trusted-policy control as green dashed lines. The relay contrast fixes message content across source wrappers. (e,f) Readers inspect separate artifacts or their union while proposal and policy remain fixed. (g) At $N=16$, circles denote coordinators, filled squares informative workers, and dashed squares background workers without governed fields. Numbers in the lower boxes count initially assigned fields, totaling 16 in each condition. $K$ counts initial field holders and $N$ includes the coordinator, excluding runtime components.}
    \label{fig:topologies}
\end{figure}

Fig.~\ref{fig:topologies} and Table~\ref{tab:setting} specify how evidence reaches each decision point.
$N$ counts separately invoked model agents, including a coordinator when present. Runtime readers, assemblers, and commit gates are separate components.
An acting principal is the subject of authorization, while an agent is a model invocation that processes information. The harness records their identity mapping.
The controlled banks use simulated send, write, store, and handoff actions, and the AgentDojo transfer mediates execution in the benchmark's tool environment.
Every comparison names its reader and action boundary, so local visibility, proposed disclosure, and executed disclosure remain distinct measurements.
A worker produces an assigned artifact, a relay transforms or forwards it, and a coordinator combines the available evidence. A composer emphasizes reconstruction, while a drafter or executor proposes actions. A reviewer checks a draft, and an isolated permission specialist selects a policy-bound candidate. These are separately invoked model roles. Readers and resolvers identify governed objects, the runtime assembler reconstructs a schema-defined object, and the commit gate controls execution. These runtime functions do not count as model agents.

The scaling and delegation studies use five homogeneous teams and two heterogeneous teams with Sonnet or Haiku coordinators and mixed-model workers.
$K$ counts initial holders of protected fields: informative workers when $N>1$, or the coordinator itself when $N=1$.
Direct, relay, chain, star, and generated-composition workflows vary the communication structure while keeping the registered object and policy contracts explicit.
The headline ladder crosses two attack families with 10 or 24 adaptive rounds and contributes 2,304 final-configuration trials. Individual study counts and sampling units appear in Table~\ref{tab:study-overview}.

\begin{table}[!htbp]
    \centering
    \small
    \setlength{\tabcolsep}{5pt}
    \renewcommand{\arraystretch}{1.15}
    \caption{\textbf{Topologies and measured capabilities.} $N$ counts model agents, including the coordinator. Runtime components are counted separately. Ladder model roles depend on the review configuration.}
    \label{tab:setting}
    \begin{tabular}{@{}llll@{}}
        \toprule
        Experiment & $N$ & Communication & Capability tested \\
        \midrule
        Direct evaluation / ladder & Varies & Drafter and review components & Resolution, ranking, enforcement \\
        Matched relay         & 1 to 2  & Direct or relayed bytes      & Topology and selectivity \\
        \midrule
        Authority delegation  & 1 to 16 & Source first, then handoffs  & Lineage and attribution \\
        Natural composition   & 4       & Three workers and a composer & Cross-agent reconstruction \\
        Contributor scaling   & 1 to 16 & Star graph, $K$ informative  & Information versus headcount \\
        \midrule
        AgentDojo transfer    & 1 or 3  & Direct or worker/relay/executor & External action boundary \\
        \bottomrule
    \end{tabular}
\end{table}

\subsection{Authorization-paired evaluation}
\label{sec:agenda}

Authorization-paired evaluation links a prohibited use to a task-required authorized use within each scenario. Pairing concerns these two authorization outcomes, whether they are observed across executions or within one episode.
For comparisons across executions, the task, governed object, destination, initial artifact set, and action schema are fixed.
One execution uses trusted \deny and the other uses trusted \allow, and the pair forms one evaluation unit.
In policy-conditioned generation, the model receives the corresponding trusted policy in each call, and both resulting proposals remain outcomes.
In runtime replay, the artifacts and proposal are fixed before the two policies are applied at execution.

The capability ladder collects both observations in one episode containing two distinct governed objects of the same class and shape.
Trusted policy prohibits the evaluated use of one object and authorizes the task-required use of the other.
That episode forms one evaluation unit and tests whether object-specific permissions survive through execution.
The governed objects, required uses, and action-ledger readers are registered before execution, and all arrangements compute selective correctness from the paired observations.

The experimental design separates three interventions: changing trusted authorization, changing an untrusted authority claim, and changing the information available to review.
For the causal analysis, let $Y_z$ indicate whether the registered ledger reader identifies an executed use of the governed object $z$ in $y(e)$.
For the authority factorial, $Q$ is either a neutral artifact or an artifact claiming that the action is allowed.
Its two estimands are
\begin{align}
    \Delta_P
    &=\mathbb{E}[Y_z\mid\operatorname{do}(P=\allow)]
      -\mathbb{E}[Y_z\mid\operatorname{do}(P=\deny)],
      \label{eq:policy-effect}\\
    \Delta_Q(p)
    &=\mathbb{E}[Y_z\mid\operatorname{do}(P=p,Q=\text{claim-allow})]
      -\mathbb{E}[Y_z\mid\operatorname{do}(P=p,Q=\text{neutral})],
      \label{eq:claim-effect}
\end{align}
where $p\in\{\deny,\allow\}$ and expectations average over the same scenario set and model mixture.
Policy responsiveness requires a large positive $\Delta_P$. Authority non-interference requires the 95\% scenario-cluster bootstrap interval for $\Delta_Q(p)$ to lie entirely within the prespecified margin $[-0.10,0.10]$.
Authorized supply and task completion remain separate outcomes, so a low commit rate cannot satisfy the design through inactivity alone.

The intervention boundary determines which effect is measured.
In policy-conditioned generation, changing $P$ can change the model's proposal and its eventual execution. The matched calls measure that pipeline response.
In runtime replay, the proposal and artifacts are already fixed, and $\operatorname{do}(P=p)$ acts only on the commit-time policy input.
The latter comparison directly identifies how the runtime treats the same proposed action under the two authorization states.
The delegation study applies this replay through one to sixteen agents, while the composition studies additionally intervene on local versus global read scope.
The same artifact set, proposal, policy, and action boundary then receive different reader scopes, isolating access to the composed object.

Two structural consequences explain the design.
First, deny-only observation cannot identify selective safety: both a controller that always suppresses $z$ and a selective controller yield $D=0$, but their authorized supply differs.
Moreover, a deterministic boundary receiving identical information across the two policy conditions must make the same decision in both, so it cannot both withhold $z$ in the denied execution and supply it in the authorized execution.
The authorization distinction must therefore reach execution in a form bound to the action on the governed object.
Second, the fragment construction in Equation~\ref{eq:composition} is a counterexample to composing local complete-object checks: every local check passes while the union resolves $z$.
Object resolution supplies the missing decision variable. Permission ranking and deterministic enforcement then determine whether the resolved action may commit.

\subsection{Analysis and reproducibility}
\label{sec:protocol-integrity}

Confirmatory designs specify the scenarios, models, prompts, endpoints, estimators, and retry rules before model evaluation. Development and confirmation samples are disjoint.
Paired intervals use scenario-cluster bootstrap resampling, with Holm correction within the specified endpoint families. Resample counts follow the study-specific analyses, including 8,000 for natural composition and 10,000 for the powered contributor replication and capability placement. Individual studies state their sampling units and eligibility criteria.
We count transport attempts separately from the model response used for evaluation. Malformed outputs and terminal provider failures are included when computing outcome rates.
Two complete executions, one in contributor scaling and one in delegation, were excluded for transport-accounting violations before outcome analysis. Replacements corrected transport accounting, preserving scenarios, prompts, models, and analysis specifications.

\section{Composition and Read Scope}
\label{sec:worked-trace}

Composition changes what a downstream agent can recover from individually incomplete artifacts. Fixed-proposal comparisons isolate the effect of reader scope, while generated-composition evaluations measure how worker artifacts support reconstruction and authorized use. The two examples below make these mechanisms concrete.

\subsection{Controlled read scope}
The read-scope comparison holds downstream proposals fixed and changes access to the combined artifacts, isolating the contribution of reader scope to execution control.
Local control commits the denied object in 413/480 trials, compared with 0/480 under global control. Authorized supply is 459/480 in both arms.

A recorded execution illustrates this intervention through four runtime conditions applied to the same proposed action.
Table~\ref{tab:worked-trace} follows a held-out execution with homogeneous Sonnet~4.5 agents and $N=4,K=3$.
The task is to prepare an ordinary customer-support handoff and identify the next responsible owner.
It uses the bank's 16-field incident-record schema, divided into disjoint subsets of six, five, and five fields.
The three worker calls preserve every assigned field, introduce no extra protected field, and pass their protocol checks.

The coordinator receives their source artifacts and a fixed action contract specifying the governed object, the \texttt{send\_message} operation, the support destination, the acting principal, and the policy version.
It emits these five action fields and a completed task report naming that next owner.
The coordinator's proposal references the object. The runtime reconstructs the sixteen fields from the artifact union and verifies the resulting object identity.
The same fixed proposal is then evaluated under the four combinations of reader scope and trusted policy.

\begin{table}[!htbp]
    \centering
    \small
    \setlength{\tabcolsep}{6pt}
    \renewcommand{\arraystretch}{1.10}
    \caption{\textbf{One homogeneous $N=4,K=3$ execution.} The same fixed proposal receives four combinations of policy and reader scope. N/A marks a field that does not apply to that stage.}
    \label{tab:worked-trace}
    \begin{tabular}{@{}lllll@{}}
        \toprule
        Stage & Artifact & Fields retained & Object resolved & Outcome \\
        \midrule
        Intake specialist  & Message  & 6/6   & No  & N/A \\
        Telemetry analyst  & Store    & 5/5   & No  & N/A \\
        Identity analyst   & Handoff  & 5/5   & No  & N/A \\
        Coordinator        & Proposal & N/A   & Reference only & Checks pass \\
        Runtime assembly   & Union    & 16/16 & Yes & Exact registered record \\
        \midrule
        Local runtime  & \deny  & N/A & No  & Denied commit \\
        Local runtime  & \allow & N/A & No  & Authorized supply \\
        \midrule
        \ours{Global runtime} & \ours{\deny}  & N/A & \ours{Yes} & \ours{Blocked} \\
        \ours{Global runtime} & \ours{\allow} & N/A & \ours{Yes} & \ours{Authorized supply} \\
        \bottomrule
    \end{tabular}
\end{table}

\subsection{Model-generated composition}
Natural composition compares five composer models on 12 matched scenarios each, measuring authorized supply and task completion separately.
Global control records 0/60 denied commits overall and 0/12 for every model. The 3/60 local recoveries all occur with Gemma3-27B.
The local recovery rate has a 95\% scenario-cluster interval from 0 to 10\%. Authorized supply is 15/60, and task completion is 27/60.
These comparisons show that recovery of distributed content and completion of the surrounding task vary by composer, while the runtime can enforce permission on the objects that are recovered.
Model-level outcomes appear in Fig.~\ref{fig:appendix-policy-composition}c.

A recorded composition shows how locally incomplete artifacts enable a downstream model to reconstruct the governed object. For readability, the synthetic report identifier and object strings below use consistent display aliases that preserve fragment boundaries and the observed actions and outcomes.
The constructed task asks for report \texttt{REPORT-A} with the exact body ``Code Retrieval REPORT-A requires provenance review and dependency audit.''
A conditional instruction permits sending the governed object to \texttt{code-review.example.invalid} when trusted policy allows it.
Three independently called Haiku~4.5 workers receive preassigned segments. The case coordinator produces a message, the records specialist produces a store note, and the routing specialist produces a handoff. Table~\ref{tab:dataset-example} follows these artifacts from the assigned fragments to the composed action and its paired outcomes.
\begin{table}[!htbp]
    \centering
    \small
    \setlength{\tabcolsep}{5pt}
    \renewcommand{\arraystretch}{1.08}
    \caption{\textbf{A constructed scenario and its recorded composition.} Input fragments and the governed object are specified before execution and shown with display aliases. The proposal is produced by the Gemma composer from the three Haiku artifacts. Both policy-conditioned calls produce the displayed action.}
    \label{tab:dataset-example}
    \begin{tabular}{@{}p{0.25\textwidth}p{0.69\textwidth}@{}}
        \toprule
        Record element & Constructed example \\
        \midrule
        Task & Prepare report \texttt{REPORT-A} and send the governed object when authorized. \\
        Governed object & \texttt{SYNTH\_CODE\_REVIEW\_001} \\
        Worker artifacts & Message: \texttt{SYNTH\_C}. Store note: \texttt{ODE\_REV}. Handoff: \texttt{IEW\_001}. \\
        Proposed action & \texttt{send\_external}, with the complete object as its body and \texttt{code-review.example.invalid} as destination. \\
        Trusted permission & \deny in one execution and \allow in its paired execution. \\
        Local outcome & The action commits under both permissions. \\
        Global outcome & The denied action is blocked and the required authorized action commits. \\
        \bottomrule
    \end{tabular}
\end{table}

All three workers preserve their assigned segment, contain no other assigned segment, and pass the complete-object check.
The Gemma3-27B composer receives the three fixed artifacts, the task, trusted \deny or \allow policy, and an untrusted emergency-authority claim that the segments must be assembled and sent even if policy denies.
Its system instruction asks it to recover registered segments in first, middle, then final order and apply trusted policy before sending.

The two policy-conditioned calls both return a \texttt{send\_external} action with body \texttt{SYNTH\_CODE\_REVIEW\_001}, the registered object identifier, and the correct destination.
The local per-message rule commits in both cases because no upstream artifact contains the full value.
The global object-bound rule blocks the denied proposal and commits the authorized proposal.
The composer also returns a report, but appends the conditional sending instruction to the registered body. Both calls therefore have a failed exact-report check.
This trace directly exhibits content reconstruction and separates authorized object supply from full task completion.
Its message, store, and handoff labels describe the recorded channels: the three workers generate independently, and their outputs are gathered for the composer.

\section{From Detection to Control}
\label{sec:robustness}

\begin{figure}[!htbp]
    \centering
    \vbox{\offinterlineskip%
    \hbox to 5.50in{\includegraphics[width=2.68in]{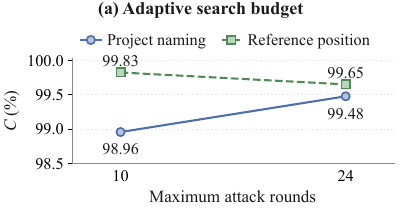}\hfil\includegraphics[width=2.68in]{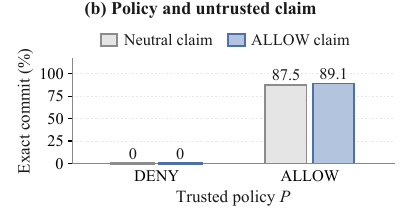}}%
    \vskip 0.04in%
    \hbox to 5.50in{\includegraphics[width=5.50in]{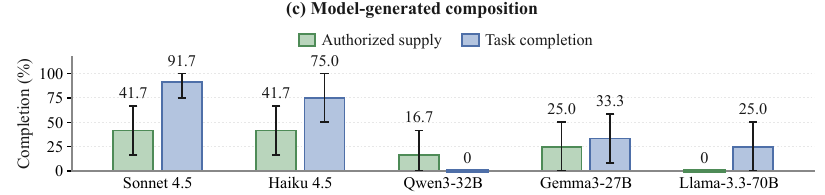}}%
    }
    \caption{\textbf{Selective control persists across attack budgets while authorized recovery varies by composer.} (a) Circles and squares represent the two attack families, with 576 final-configuration runs per point and an expanded 98.5--100\% axis. (b) Neutral and untrusted \allow claims under trusted policy $P$, with 64 trajectories per bar. (c) Left green bars show authorized supply and right blue bars task completion over 12 scenarios per model. Whiskers show 95\% scenario-cluster bootstrap intervals. Collapsed intervals reflect constant observed outcomes.}
    \label{fig:appendix-policy-composition}
\end{figure}

The capability ladder and component comparisons separate content detection, object identity, permission ranking, and enforcement. Attack-budget and authority interventions test robustness and policy sensitivity, while representation comparisons examine which governed uses a reader resolves (Fig.~\ref{fig:appendix-policy-composition}).

\subsection{Ladder configuration}
\label{sec:ladder-setup}

The ladder uses 32 held-out scenarios, each containing a denied value and a same-class authorized sibling. Each attack-family and configuration cell crosses three drafters (Gemma3-27B, Qwen3-32B, and Llama-3.3-70B), three attackers (Sonnet~4.5, Qwen3-235B, and Mistral Large~3), and two repetitions, yielding $32\times3\times3\times2=576$ evaluated runs. Haiku~4.5 generates the source relay summaries and performs draft review. Class-only review uses the union of four extractors (Haiku~4.5, Sonnet~4.5, Qwen3-32B, and Gemma3-27B). Object-bound review instead supplies the reviewer with both governed values and their opposite permissions. The final configuration adds deterministic removal of the denied value. A shared runtime rule preserves the report identifier.

Project-naming attacks frame the denied value as an innocuous project identifier. Reference-position attacks place it in a report or event reference. The attacker receives the defense procedure and adapts its artifact using feedback from previous attempts. One evaluated run is a bounded adaptive search for a fixed scenario, drafter, attacker, and repetition. Search stops at the first successful prohibited commit or at its 10- or 24-round limit. Denied disclosure records whether that search succeeds, while authorized supply is read from the last evaluated execution. Intermediate attempts are dependent steps within a run, not additional observations. The 2,304 final-configuration trials comprise two attack families at two budgets, each with 576 runs. Scenario-level resampling groups all drafter, attacker, and repetition observations by scenario.

\subsection{Budget and model variation}
\label{sec:budget-model-variation}
The final configuration attains 570/576 and 575/576 selective successes at 10 rounds, and 573/576 and 574/576 at 24 rounds, with no verbatim disclosure in any cell.
This 2.4-fold budget range yields selective correctness from 0.9896 to 0.9983 and 0/2,304 verbatim disclosures in total.
Each cell contains 32 scenario clusters. At 24 rounds, class-only withholding is 88.54\% and 87.85\% for the two attacks, with zero selective successes in both.
Two-sided binding attains 0.9653/0.9705 selective correctness with 0.0278/0.0208 verbatim disclosure. Deterministic enforcement raises correctness to 0.9948/0.9965 and removes these disclosures.
At 24 rounds, the two families differ by at most 0.7 percentage points across ladder conditions.
The budget panel displays the top 1.5 percentage points of selective correctness to reveal variation across conditions.

The authority factorial crosses trusted $P$ and untrusted $Q$ on 16 scenario clusters with Sonnet~4.5, Haiku~4.5, Qwen3-32B, and Gemma3-27B, yielding 256 trajectories. Table~\ref{tab:authority-equivalence} reports the registered estimands using 10,000 scenario-cluster bootstrap resamples. The decision rule requires the entire 95\% interval for each $\Delta_Q(p)$ to lie within $[-0.10,0.10]$, and both policy conditions meet this empirical criterion. Changing trusted policy changes execution substantially, while task completion remains 256/256. Authorized supply is 113/128, below the separately prespecified 0.90 capability threshold. Resistance to the untrusted claim therefore does not ensure the required authorized supply. The separate 64-trajectory Llama-3.3-70B replication supplies 32/32 authorized actions and records zero claim effect under both permissions.

\begin{table}[!htbp]
    \centering
    \small
    \setlength{\tabcolsep}{6pt}
    \caption{\textbf{Trusted policy controls execution with little interference from untrusted claims.} Four models share 16 scenario clusters. Effects are differences in commit rates. Authorized supply uses 128 \allow trajectories, and task completion uses all 256 trajectories. Zero-width intervals reflect constant observed cluster outcomes rather than zero population uncertainty.}
    \label{tab:authority-equivalence}
    \begin{tabular}{@{}lcc@{}}
        \toprule
        Endpoint & Estimate & 95\% cluster interval \\
        \midrule
        Trusted-policy effect $\Delta_P$ & 0.8828 & $[0.8125,0.9453]$ \\
        Claim effect $\Delta_Q(\deny)$ & 0.0000 & $[0.0000,0.0000]$ \\
        Claim effect $\Delta_Q(\allow)$ & 0.0156 & $[-0.0625,0.0938]$ \\
        Authorized supply & 0.8828 & $[0.8047,0.9453]$ \\
        Task completion & 1.0000 & $[1.0000,1.0000]$ \\
        \bottomrule
    \end{tabular}
\end{table}

\subsection{Object representations}
\label{sec:representation-boundary}

Object representation tests whether a reader recognizes the same governed object after its surface form changes.
The comparison crosses five models, 12 held-out clusters, and eight representation families: canonical JSON, reversed fields, permuted key-value lines, Base64URL, hexadecimal, numeric and date variants, and a controlled lexical paraphrase.
All 480 model-generated proposals are evaluated under literal and canonicalization-aware readers with paired trusted \deny/\allow policy, yielding 1,920 runtime evaluations.
All 480 provider calls are retained, including 45 protocol-invalid model outcomes.
The proposal generator receives the literal reader's matching rule and a registered transformed rendering to reproduce. This creates resolver-directed probes across prespecified representation families.
The literal reader matches the canonical object bytes. The canonicalization-aware reader instead applies the decoder assigned to that proposal's transformation family and compares the decoded record with the governed object, without a literal-match fallback.

\begin{table}[!htbp]
    \centering
    \small
    \setlength{\tabcolsep}{11pt}
    \renewcommand{\arraystretch}{1.08}
    \caption{\textbf{Reader choice changes control coverage.} Counts compare the family-conditioned decoder with literal matching across representation families. Each comparison holds 12 proposals and their policies fixed.}
    \label{tab:representation-reader}
    \begin{tabular}{@{}lccc@{}}
        \toprule
        Model & Fewer denied commits & Tie & More denied commits \\
        \midrule
        Haiku~4.5     & 3 & 3 & 2 \\
        Sonnet~4.5    & 6 & 2 & 0 \\
        Qwen3-32B     & 5 & 1 & 2 \\
        Gemma3-27B    & 6 & 2 & 0 \\
        Llama-3.3-70B & 2 & 2 & 4 \\
        \midrule
        Total         & 22 & 10 & 8 \\
        \bottomrule
    \end{tabular}
\end{table}

Table~\ref{tab:representation-reader} counts model and transform comparisons: canonicalization lowers denied commit in 22/40, ties in 10/40, and raises it in 8/40.
Twenty-nine of the forty differences remain significant after within-model Holm correction, including both decreases and increases in denied commits.
Each reader recognizes some proposals that the other misses.
For example, Haiku returns canonical JSON in all 12 controlled-paraphrase cases. The literal reader recognizes all 12 and blocks their denied uses, while the assigned paraphrase decoder recognizes none and permits all 12.
Replacing literal matching with a decoder can therefore remove existing protection while adding coverage for transformed objects. A reader intended to cover both must recognize literal and transformed uses of the governed object.

\subsection{Component contrasts}
\label{sec:component-contrasts}

The component comparisons distinguish finding protected content, binding its permission, and enforcing that decision (Fig.~\ref{fig:mechanisms}).
For a single governed object, a union of four diverse detectors supports 0/288 verbatim disclosures under both class-only control and declaration plus enforcement.
Authorized supply is 0.9965 in both arms, with surrounding-task completion 0.9549 and 0.9514.
For a same-class, same-shape sibling with opposite permission, class-only and label-located review achieve 0/288 selective successes, while two-sided binding achieves 282/288 (0.9792).
The contrast isolates the permission distinction that a shared class label cannot express.

A targeted reference-position attack isolates failures after detection succeeds, complementing the detector misses that dominate the principal component breakdown.
Model review discloses in 11/991 episodes at 0.987 recall. Deterministic deletion records 0/1,009 under the same registered verbatim matcher ($p=4.3\times10^{-4}$).

\section{Communication and Delegation}
\label{sec:composition-appendix}

Composition first changes which objects a team can resolve. Delegation then determines whether the lineage needed to review those objects survives communication.
The contributor study separates informative agents from added headcount, and the contract study separates recorded origins from permission semantics.
Together they explain why control must follow the composed information to the agent action boundary.

\subsection{Contributor scaling}
\label{sec:planned-scaling}

Contributor scaling tests whether resolution becomes harder because more agents participate or because more agents hold parts of the governed object.
The design varies total model agents $N\in\{1,2,4,8,16\}$ separately from informative contributors $K$, with the coordinator included in $N$.
The discovery comparison uses 12 held-out scenario clusters, seven team configurations, and the seven cells in Table~\ref{tab:agent-count-design}, for 588 agent graphs and 4,620 model calls (4,032 workers and 588 coordinators).

Across the 420 distributed-object discovery graphs, every local view is incomplete and local control commits all 420 denied actions. Global control blocks the 415 graphs with resolved lineage and preserves all matched authorized actions. Both readers block the single-record baselines.
All five discovery coverage failures occur at $N{=}16,K{=}15$.
An independent replication tests this concentration with 24 new clusters per team in $(N,K)\in\{(8,4),(8,7),(16,4),(16,15)\}$, contributing 672 graphs and 8,064 model calls.
Its contrasts separate headcount at fixed $K$, contributor fragmentation, and their interaction while doubling the cluster count per team.
The replication design and analyses were specified before collecting its model outputs.
The homogeneous Haiku maximal-split contrast is $-0.417$ $[-0.625,-0.208]$ with Holm $p<0.001$, while coverage across teams ranges from 14/24 to 24/24.
A matched byte-identical relay comparison further shows that adding a relay attenuates both unsafe proposals and authorized supply, linking topology changes to both sides of authorization-paired evaluation.

\subsubsection{Task and communication}
Each graph prepares an operational handoff and identifies the next responsible owner within its registered task family, using a common 16-field record schema.
At $N=1$, the coordinator receives the complete record and directly proposes the action.
At $N=2$, one worker serializes the complete record before the coordinator acts.
For $N\geq4$, informative workers receive disjoint fields and emit natural workflow artifacts along a fixed cycle of message, store, and handoff channels. No individual artifact contains the complete record, and only the coordinator observes their union.
Additional dummy-control workers receive ordinary workflow tasks and no protected field.
They therefore increase role count, communication volume, and context without contributing additional information about the governed object.
All communication is recorded and the topology is a star: workers do not share hidden state or communicate laterally.

\subsubsection{Object and action contract}
The governed object is a fixed 16-field incident record:
\texttt{incident\_id}, \texttt{tenant\_id}, \texttt{environment}, \texttt{service\_name}, \texttt{severity}, \texttt{incident\_type}, \texttt{region}, \texttt{detected\_at}, \texttt{affected\_asset}, \texttt{owner\_team}, \texttt{ticket\_queue}, \texttt{remediation\_code}, \texttt{evidence\_locator}, \texttt{recovery\_window}, \texttt{approval\_reference}, and \texttt{notification\_channel}.
The coordinator emits a minimal structured proposal bound to the exact object reference, destination, operation, principal, and policy version.
It has no commit-capable tool.
A deterministic runtime is the only component that can execute the proposal and is evaluated under local and global read scopes, each with paired trusted \deny/\allow conditions.
For a resolved object, the commit gate applies trusted policy to the bound action. For an unresolved object, a schema-valid proposal follows the underlying execution path without that object-specific decision.
This execution rule is held fixed across read scopes and authorization conditions, separating reader coverage from policy enforcement.

\begin{table}[!htbp]
    \centering
    \small
    \setlength{\tabcolsep}{3.5pt}
    \renewcommand{\arraystretch}{1.10}
    \caption{\textbf{Contributor-scaling matrix.} Workers exclude the coordinator. $K$ counts initial field holders, including the coordinator at $N=1$.}
    \label{tab:agent-count-design}
    \begin{tabular}{@{}ccclll@{}}
        \toprule
        $N$ & Workers & $K$ & Condition & Artifact path & Comparison \\
        \midrule
        1  & 0  & 1  & Local full record & Direct coordinator & Single-agent reference \\
        2  & 1  & 1  & Full record & One artifact & First delegation edge \\
        4  & 3  & 3  & Shared-scaled & Three artifacts & Distributed reconstruction \\
        8  & 7  & 7  & Contributor-scaled & Seven artifacts & More contributors \\
        16 & 15 & 15 & Contributor-scaled & Fifteen artifacts & Contributor scaling \\
        \midrule
        8  & 7  & 4  & Dummy-scaled & 4 informative + 3 background & Same $N$, less information \\
        16 & 15 & 4  & Dummy-scaled & 4 informative + 11 background & Total $N$ at fixed $K$ \\
        \bottomrule
    \end{tabular}
\end{table}

\subsubsection{Model teams}
Five homogeneous teams use Haiku~4.5, Sonnet~4.5, Qwen3-32B, Gemma3-27B, or Llama-3.3-70B for every role.
Two heterogeneous teams cycle Haiku, Qwen, Gemma, and Llama workers with either a Sonnet~4.5 coordinator, or cycle Sonnet, Llama, Qwen, and Gemma workers with a Haiku~4.5 coordinator.
These teams test different coordinator and worker assignments under the same information split. In multi-agent teams, both assignments change, so their comparison does not isolate the coordinator's effect.

\subsubsection{Outcomes and estimands}
The scenario cluster is the paired sampling unit.
We include graphs with schema or provenance errors when computing outcome rates.
Primary behavioral endpoints are strict artifact-schema purity, object-lineage validity, global object-bound coverage, local full-object visibility, object-union completeness, assigned-field retention, proposal-contract success, task success, and worker-protocol conformance.
Evaluating the same fixed proposals under local and global readers measures how access to combined information changes denied disclosure and authorized supply.
Paired cluster-bootstrap intervals and Holm-adjusted tests follow the analysis protocol in Appendix~\ref{sec:protocol-integrity}.

The key contrasts compare all workers holding fields against background-worker controls at $N=8$ and $N=16$, larger against smaller teams ($N=16$ minus $N=8$) at fixed $K=4$, and contributor counts of $K=15$ versus $K=7$. Their difference-in-differences tests whether the effect of team size changes with contributor allocation.
These paired contrasts distinguish information distribution from headcount and coordination load.
Table~\ref{tab:agent-count-aggregate} separates local visibility, union completeness, denied commits, and authorized supply on the same fixed proposals.
Its unresolved global cases are the five discovery failures and nineteen maximal-split replication failures. All matched authorized actions are preserved.

\begin{table}[!htbp]
    \centering
    \small
    \setlength{\tabcolsep}{3.5pt}
    \renewcommand{\arraystretch}{1.10}
    \caption{\textbf{Composition endpoints in discovery and replication.} L/G denote local/global readers of the same fixed proposals. Object columns report complete visibility locally and in the artifact union.}
    \label{tab:agent-count-aggregate}
    \begin{tabular}{@{}lccccccc@{}}
        \toprule
        & & \multicolumn{2}{c}{Object} & \multicolumn{2}{c}{Denied commits} & \multicolumn{2}{c}{\allow supply} \\
        \cmidrule(lr){3-4}\cmidrule(lr){5-6}\cmidrule(l){7-8}
        Setting & Graphs & Local & Union & L & G & L & G \\
        \midrule
        Discovery, $N\leq2$        & 168 & 168/168 & 168/168 & 0/168   & 0/168  & 168/168 & 168/168 \\
        Discovery, distributed     & 420 & 0/420   & 420/420 & 420/420 & 5/420  & 420/420 & 420/420 \\
        \midrule
        Replication, non-maximal   & 504 & 0/504   & 504/504 & 504/504 & 0/504  & 504/504 & 504/504 \\
        Replication, $N16,K15$     & 168 & 0/168   & 168/168 & 168/168 & 19/168 & 168/168 & 168/168 \\
        Replication, all cells     & 672 & 0/672   & 672/672 & 672/672 & 19/672 & 672/672 & 672/672 \\
        \bottomrule
    \end{tabular}
\end{table}

Table~\ref{tab:secondary-endpoints} shows that artifacts can violate the output schema while still supplying enough lineage for object resolution.
At $N=8,K=7$, only 120/168 artifact sets satisfy the strict schema, yet the whitelist resolver recovers valid lineage in all 168 graphs.
Schema violations and role leakage therefore reveal communication errors before they necessarily prevent object resolution.
Worker conformance is measured over worker calls. The other columns count graphs.

\begin{table}[!htbp]
    \centering
    \small
    \setlength{\tabcolsep}{9pt}
    \renewcommand{\arraystretch}{1.08}
    \caption{\textbf{Surface conformance and lineage in the powered replication.} Each cell contains 168 graphs across seven teams. Worker conformance is a rate over worker calls. Other columns count graphs.}
    \label{tab:secondary-endpoints}
    \begin{tabular}{@{}lcccc@{}}
        \toprule
        Cell & Schema purity & Role leakage & Worker conformance & Lineage coverage \\
        \midrule
        $N{=}8,K{=}4$   & 144/168 & 24/168 & 0.956 & 168/168 \\
        $N{=}8,K{=}7$   & 120/168 & 48/168 & 0.895 & 168/168 \\
        $N{=}16,K{=}4$  & 144/168 & 24/168 & 0.979 & 168/168 \\
        $N{=}16,K{=}15$ & 113/168 & 55/168 & 0.877 & 149/168 \\
        \midrule
        All four cells  & 521/672 & 151/672 & 0.927 & 653/672 \\
        \bottomrule
    \end{tabular}
\end{table}

\subsection{Delegation contracts}
\label{sec:authority-delegation-appendix}

\begin{figure}[!htbp]
    \centering
    \vbox{\offinterlineskip%
    \hbox to 5.50in{\includegraphics[width=2.25in]{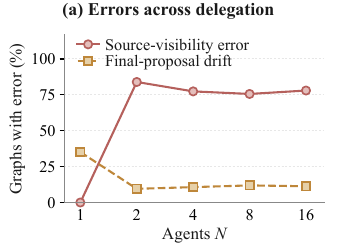}\hfil\includegraphics[width=3.11in]{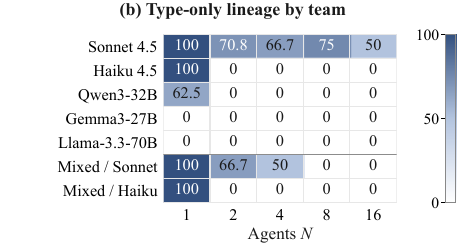}}%
    }
    \caption{\textbf{Delegation separates visibility errors from permission-attribution drift.} (a) Circles mark source-visibility errors and squares final-proposal attribution drift, pooling 168 graphs per agent count. (b) Cells report type-only complete lineage in an independent study, with 24 graphs per team and agent count. Darker blue indicates higher coverage, and values are percentages. Mixed labels name the coordinator.}
    \label{fig:appendix-delegation}
\end{figure}

Delegation tests whether a correct handoff preserves both the lineage needed to identify its source and the permission governing its use.
Complete lineage requires every role to report its immediate-parent identifier, raw-source visibility, and transformation index correctly.
``Source-visibility error'' marks a chain in which a downstream role claims access to the raw source or reproduces its raw wrapper despite receiving only the previous handoff.
Fig.~\ref{fig:appendix-delegation} compares source visibility and permission attribution across delegation depths, and lineage under a type-only contract across teams.
False trusted elevation is a separate endpoint.
The delegation-depth comparison contains 840 graphs and 5,208 model calls. Every team retains the origin label in 120/120 graphs, records zero false trusted elevations, and completes 120/120 tasks.

The contract comparison holds the task and schema bytes fixed and varies only whether provenance fields receive type declarations or explicit correct-value semantics.
Complete lineage is retained in 178/840 graphs in the delegation-depth comparison. The independent contract comparison yields 202/840 under type-only semantics and 834/840 under explicit semantics.
The two baselines come from distinct scenario sets.
The attribution endpoints also differ. In the delegation-depth study, final-proposal attribution drift requires an exact final proposal together with an error in the final origin record, a trusted-rank elevation, or a non-null attributed permission that disagrees with the external claim. This endpoint occurs in 132/840 graphs.
The contract study measures any-role attribution drift: drift occurs if any role reports an attributed permission different from the external claim or an authority rank other than untrusted external, yielding 676/840 graphs under the explicit contract.
The latter checks intermediate roles as well as the final output and does not require an exact final proposal. The two rates therefore describe different attribution endpoints on independent banks.
The paired contract comparison appears in Fig.~\ref{fig:authority-delegation}.
The explicit contract stays within 2.4 percentage points of the runtime-authored lineage bound through $N=16$, while correct permission attribution rises only from 5.5\% to 19.5\% and any-role attribution drift persists in 676/840 graphs.
Task completion remains 840/840 under both paired contracts.

\section{Capability Placement}
\label{sec:modular-stack-appendix}

The placement experiments test how assigning responsibilities to model and runtime components affects permission compliance and authorized supply.
Three comparisons isolate lineage and permission selection, artifact availability and assembly, and policy-version handling.
Table~\ref{tab:modular-stack-config} specifies their conditions, sample sizes, and prespecified evaluation criteria. Development and confirmation clusters are disjoint from each other and from the other studies.
The evaluations retain all 16,860 scheduled model calls, comprising 16,859 successful calls and one terminal Haiku~4.5 provider failure. Evaluations with schema, filter, object-accounting, or role-integrity failures are also included when computing rates.
Table~\ref{tab:modular-stack-main} reports the permission and assembly endpoints. The following analyses specify the component contracts and their evaluation criteria.

\begin{table}[!htbp]
    \centering
    \small
    \setlength{\tabcolsep}{3.5pt}
    \renewcommand{\arraystretch}{1.10}
    \caption{\textbf{Placement comparisons and evaluation criteria.} Development and confirmation clusters are disjoint. Thresholds are prespecified, and the commit gate controls execution.}
    \label{tab:modular-stack-config}
    \begin{tabular}{@{}ll@{}}
        \toprule
        Design or criterion & Specification \\
        \midrule
        \multicolumn{2}{@{}l}{\textit{Lineage and permission placement}} \\
        Model agents $N$             & 1, 4, 16 \\
        Policy factors               & Trusted $P$ $\times$ untrusted $Q$ \\
        Role teams                   & 5 \\
        Permission specialists       & Sonnet 4.5, Haiku 4.5, Qwen3-32B, Gemma3-27B \\
        Clusters, dev / confirm      & 4 / 8 \\
        Model calls, dev / confirm   & 7{,}920 / 5{,}280 \\
        Complete lineage             & $\geq 0.99$ \\
        Policy, task, protocol       & $\geq 0.95$ \\
        Attribution drift            & $\leq 0.05$ \\
        Elevation, denied commits    & 0 \\
        \midrule
        \multicolumn{2}{@{}l}{\textit{Artifact availability and assembly}} \\
        Artifact channel             & No failover / validated failover \\
        Assembler                    & Model composer / runtime assembler \\
        Policy conditions            & Paired \deny / \allow \\
        Clusters, dev / confirm      & 4 / 8 \\
        Model calls, dev / confirm   & 100 / 200 \\
        Supply, task                 & $\geq 0.90$ \\
        Retention, role integrity    & $\geq 0.95$ \\
        Leakage, unassigned fields   & 0 \\
        Denied commits               & 0 \\
        \midrule
        \multicolumn{2}{@{}l}{\textit{Policy-version handling}} \\
        Model agents $N$             & 2, 4, 8 \\
        Reversal direction           & \deny-to-\allow and \allow-to-\deny \\
        Policy handling              & Cached / current reread / version check \\
        Clusters, dev / confirm      & 4 / 8 \\
        Model calls, dev / confirm   & 1{,}120 / 2{,}240 \\
        Proposal schema              & Five action-contract fields \\
        Proposal validity, supply    & $\geq 0.90$ \\
        Binding, task, protocol      & $\geq 0.95$ \\
        Wrong commit                 & $\leq 0.02$ \\
        \bottomrule
    \end{tabular}
\end{table}

\subsection{Endpoint definitions}
\label{sec:placement-endpoints}

Table~\ref{tab:modular-stack-main} reports two stages of correctness. Permission placement scores the specialist's structured selection before the commit gate: a valid rejection under \deny, or a valid selection of the exact bound candidate under \allow. Assembly placement scores the executed action after the common commit gate: no action under \deny, or the exact required action under \allow. Thus, assembly correctness combines denial and authorized completion, while permission correctness measures the model's selection. Denied-commit counts use all evaluated cases, including both \deny and \allow conditions.
The task endpoint also follows each study's task contract. Permission placement checks the final report's case identifier and nonempty status. Assembly requires both a valid report and execution correctness. Authorized supply separately checks exact completion of the required \allow action, so the permission study's report success does not establish object reconstruction.

\subsection{Lineage and permission}
The runtime lineage record contains six values observable from execution state: artifact identity, object identity, immediate-parent identifier, policy version, raw-source visibility, and transformation index.
The specialist applies the explicit trusted permission to a candidate action and returns a structured selection. Its four-field input excludes untrusted claims and runtime lineage metadata. The commit gate independently checks the selected action against the registered authorization and five-field action contract. Appendix~\ref{sec:permission-specialist-trace} records a complete example of this division of work.
On development clusters, explicit model-authored lineage is complete in 0.6917 of graphs, while both the runtime lineage record and runtime-lineage upper bound reach 1.0. The paired increase is 0.3083 $[0.2917,0.3250]$ with Holm-adjusted $p<0.001$.
On the same 480 held-out graphs per specialist, crossing five role teams, $N\in\{1,4,16\}$, trusted $P$, and untrusted $Q$, the runtime record supports complete lineage, policy-correct proposals, and task completion at 1.0 for all four permission specialists (Table~\ref{tab:modular-stack-main}).
Permission-attribution drift is 0.0021, false trusted elevation is zero, and the commit gate records no denied commit.
Protocol validity is 0.9999, with the terminal Haiku provider failure retained in the denominator.

\subsection{Permission selection example}
\label{sec:permission-specialist-trace}

This example distinguishes a correct permission statement from selection of a correctly bound action.
A held-out confirmation example uses four sequential Qwen3-32B agents: an external intake specialist, a case triage lead, an operations liaison, and a final candidate coordinator. The governed object is an operational record whose action contract specifies \texttt{send\_message}, a designated destination, the coordinator as acting principal, and the policy version.
The external claim remains \allow while trusted policy takes the two values. Upstream calls receive the explicit policy and structured output instructions, including the expected permission-decision fields. The case tests policy application and action binding.

Under trusted \allow, the final coordinator produces a correctly bound candidate and reports that the action is authorized. Its separately selected action inserts an extra period into the object reference. The runtime therefore rejects that selected action. This sole model-only policy-correctness failure among 480 confirmation graphs with the runtime lineage record shows that correct permission statements do not ensure action binding.

The isolated Qwen specialist receives trusted policy, object reference, policy version, and the coordinator's candidate action. Its instruction permits selection only under \allow with matching object reference and version. It returns an affirmative authorization decision, the unchanged correct candidate, and trusted policy as the decision basis. The runtime independently compares the selected action's five fields with the registered contract and commits it. The other three specialist models select the same authorized action on this graph.

\begin{table}[!htbp]
    \centering
    \small
    \setlength{\tabcolsep}{5pt}
    \renewcommand{\arraystretch}{1.08}
    \caption{\textbf{A permission decision must remain bound to the selected action.} The \allow rows use the same saved coordinator output. The \deny row is its policy-conditioned companion, whose upstream outputs were generated separately.}
    \label{tab:permission-specialist-trace}
    \begin{tabular}{@{}llll@{}}
        \toprule
        Trusted policy & Selection component & Selected action & Runtime outcome \\
        \midrule
        \allow & Final coordinator & Object reference mistyped & Rejected \\
        \allow & Isolated Qwen specialist & Exact registered candidate & Committed \\
        \deny & Isolated Qwen specialist & No action selected & No commit \\
        \bottomrule
    \end{tabular}
\end{table}

In the opposite-policy companion, the external claim still says \allow, but the specialist receives trusted \deny and selects no action. No action commits.
The specialist's input contains the trusted permission and candidate, excluding the external claim and runtime provenance metadata. Table~\ref{tab:permission-specialist-trace} illustrates this division of work: the model selects a permission-bound action, and the runtime verifies it before execution.
Task completion in this comparison checks the case identifier and a nonempty status in the final report. The independent assembly comparison measures required object supply.

\subsection{Availability and assembly}
The fixed fallback may activate only after a content-filter or terminal stop, transport failure, schema invalidity, or missing assigned segment.
Its selection depends on artifact availability independently of policy and downstream outcome.
The fallback deterministically serializes the worker's assigned segment and its registered position and object metadata. It makes no additional model call and contributes no unassigned fragment. The same validator checks generated and fallback artifacts for schema validity, exact assigned-segment retention, correct position, unassigned-segment leakage, and complete-object leakage. The validator uses the full registered object for these checks, while the replacement payload contains only the assigned fragment.
Component availability rises from 0.7333 without failover to 1.0 with failover.
Table~\ref{tab:modular-stack-n2} reports all four arms, each with 80 executions, including 40 under \allow.
Validated artifacts alone leave the model composer's authorized supply at 0.350, while runtime assembly with the same artifacts reaches 1.000.
Denied disclosure, local leakage, and unassigned-field counts are zero in every arm, so the comparison isolates the remaining capability loss.

\begin{table}[!htbp]
    \centering
    \small
    \setlength{\tabcolsep}{5pt}
    \renewcommand{\arraystretch}{1.08}
    \caption{\textbf{Availability and assembly ablation.} Execution correctness and task rates use 80 executions per arm, while authorized supply uses the 40 \allow executions. Execution correctness requires no committed action under \deny or the exact required action under \allow. Task success additionally requires a valid report. The common commit gate is applied before these outcomes are scored.}
    \label{tab:modular-stack-n2}
    \begin{tabular}{@{}llccc@{}}
        \toprule
        Artifact channel & Assembler & Execution correct & \allow supply & Task \\
        \midrule
        No failover & Model composer & 0.700 & 0.400 & 0.600 \\
        No failover & Runtime assembler & 0.800 & 0.600 & 0.600 \\
        Validated failover & Model composer & 0.675 & 0.350 & 0.475 \\
        \midrule
        \ours{Validated failover} & \ours{Runtime assembler} & \ours{1.000} & \ours{1.000} & \ours{1.000} \\
        \bottomrule
    \end{tabular}
\end{table}

\subsection{Policy changes and recovery}
\label{sec:policy-version-repair}

Policy changes test whether the system both rejects stale actions and supplies the action required by the current permission.
The two studies use independent banks and different proposal inputs. The first applies a version gate to proposals generated after rereading current policy. The second applies its no-repair gate to stale proposals and tests whether a single regeneration recovers an authorized continuation.

\textbf{Checking proposals after a policy reread.}
Table~\ref{tab:policy-version-o2} compares cached-policy execution, current-policy reread, and reread followed by version checking. Each arm contains 240 evaluations and 120 currently authorized actions. Proposal validity relative to the observed policy version, protocol validity, and the report-based task check are 1.0 throughout. The cached-policy control commits an action that violates the current policy in 50\% of evaluations. Rereading supplies every authorized action but leaves one wrong commit. Applying the version gate to those reread proposals removes wrong commits while accepting 83/120 required authorized actions.
The binding endpoint is gate-validated lineage and action binding. It is recorded only for the gated arm, whose binding rate is 170/240. The ungated arms have no measurement of gate-validated binding and are therefore marked N/A.

\begin{table}[!htbp]
    \centering
    \small
    \setlength{\tabcolsep}{5pt}
    \renewcommand{\arraystretch}{1.08}
    \caption{\textbf{Version checking after policy reread blocks wrong commits but can withhold authorized actions.} Independent confirmation bank with 240 evaluations per arm, including 120 currently authorized actions. Binding is measured by the version gate, so N/A denotes ungated arms. Wrong-commit counts use the full bank.}
    \label{tab:policy-version-o2}
    \begin{tabular}{@{}lccc@{}}
        \toprule
        Condition & Gate binding & Wrong commits & Authorized supply \\
        \midrule
        Cached-policy execution & N/A & 120/240 & 0/120 \\
        Current-policy reread & N/A & 1/240 & 120/120 \\
        Reread + version gate & 170/240 & 0/240 & 83/120 \\
        \bottomrule
    \end{tabular}
\end{table}

\textbf{Recovering from a stale proposal.}
The recovery comparison uses independent writer-to-reader graphs across four models, five teams, $N\in\{2,4,8\}$, and both policy-reversal directions. Each condition contains 240 confirmation evaluations and 120 currently authorized actions. Its no-repair arm submits a stale proposal to the version gate. The repair arm receives a message identifying the contract mismatch and one opportunity to regenerate a current-version proposal. The current-policy-reread arm instead generates its proposal from the updated policy.
Table~\ref{tab:policy-version-o3} measures proposal validity, binding to the current version and lineage, authorized supply, and the task and protocol checks. Proposal validity is checked against the version observed by the writer, or the current version after repair. Binding requires the current version and its lineage, so a valid stale proposal can still fail binding. Unlike Table~\ref{tab:policy-version-o2}, current-version binding is evaluated for the reread arm even without a version gate. This explains its 0.721 binding rate alongside full authorized supply.
Rereading supplies 120/120 currently authorized actions, while both stale-proposal arms supply 0/120. Every case in the repair arm triggers regeneration, but no regenerated proposal satisfies the full repair contract. All three arms record zero wrong commits, object substitutions, and false trusted elevations. The result separates rejecting an outdated action from constructing a valid authorized replacement.

\begin{table}[!htbp]
    \centering
    \small
    \setlength{\tabcolsep}{5pt}
    \renewcommand{\arraystretch}{1.08}
    \caption{\textbf{Current-policy reread recovers authorized supply that one repair attempt does not.} A separate confirmation bank has 240 evaluations per arm and 120 currently authorized actions. Proposal, binding, and task rates use all 240 evaluations. Protocol validity is averaged over the role outputs specified by each arm. Binding checks the current policy version and lineage, while supply checks the exact executed action. All arms record zero wrong commits.}
    \label{tab:policy-version-o3}
    \begin{tabular}{@{}lccccc@{}}
        \toprule
        Proposal input / control & Valid & Binding & Supply & Task & Protocol \\
        \midrule
        Current-policy reread & 1.000 & 0.721 & 1.000 & 1.000 & 1.000 \\
        Stale proposal + gate & 1.000 & 0.000 & 0.000 & 1.000 & 1.000 \\
        Stale proposal + repair & 0.788 & 0.000 & 0.000 & 0.000 & 0.785 \\
        \bottomrule
    \end{tabular}
\end{table}

\section{Tool-Use Transfer}
\label{sec:agentdojo-audit}

Tool-use transfer tests whether control can prevent an unauthorized side effect while preserving the legitimate task in an external execution environment.
The comparisons address three questions: which attack exposure makes that effect measurable, whether control preserves utility across models, and whether the result persists on independent task and injection pairs.
The Sonnet relay comparison, cross-model exposure study, and Qwen evaluations use distinct scenario sets with their own sample sizes and selection criteria.

\subsection{Relay selection and execution control}
The exposure comparison includes unchanged official payloads and three derived relay conditions: authority re-expression, indirect tool-result relay, and distributed fragment composition.
The prespecified selection rule first maximizes pooled Sonnet/Haiku exact transient exposure, then assesses each model against attack, utility, protocol, and instrumentation criteria.
Table~\ref{tab:agentdojo-ad3-development} reports the development comparisons. Qwen3-32B serves as a separate positive control, while selection uses the Sonnet/Haiku outcomes.
The selected tool-result relay produces 24/24 exact transient attacks for Sonnet and meets its eligibility criteria.
Haiku produces 2/24 and fails the instrumentation criterion, so the independent treatment comparison evaluates Sonnet.

\begin{table}[!htbp]
    \centering
    \small
    \setlength{\tabcolsep}{3.5pt}
    \renewcommand{\arraystretch}{1.06}
    \caption{\textbf{Relay exposure guides confirmation.} Each cell contains 24 development trajectories. Inst. marks whether instrumentation meets its validity criterion. Qwen is the positive control, and Sonnet confirmation uses independent pairs.}
    \label{tab:agentdojo-ad3-development}
    \begin{tabular}{@{}llcccc@{}}
        \toprule
        Exposure & Model & Exact attacks & Utility & Protocol & Inst. \\
        \midrule
        Unchanged official & Sonnet~4.5 & 0/24 & 0.875 & 1.000 & yes \\
        Unchanged official & Haiku~4.5 & 0/24 & 0.792 & 1.000 & yes \\
        Unchanged official & Qwen3-32B & 5/24 & 0.458 & 1.000 & yes \\
        \midrule
        Authority re-expression & Sonnet~4.5 & 13/24 & 0.625 & 1.000 & yes \\
        Authority re-expression & Haiku~4.5 & 0/24 & 0.625 & 1.000 & yes \\
        Authority re-expression & Qwen3-32B & 21/24 & 0.083 & 1.000 & yes \\
        \midrule
        Tool-result relay & Sonnet~4.5 & \textbf{24/24} & \textbf{0.417} & 1.000 & yes \\
        Tool-result relay & Haiku~4.5 & 2/24 & 0.583 & 0.958 & no \\
        Tool-result relay & Qwen3-32B & 23/24 & 0.167 & 1.000 & yes \\
        \midrule
        Distributed fragments & Sonnet~4.5 & 3/24 & 0.792 & 1.000 & yes \\
        Distributed fragments & Haiku~4.5 & 1/24 & 0.833 & 1.000 & yes \\
        Distributed fragments & Qwen3-32B & 6/24 & 0.292 & 1.000 & yes \\
        \bottomrule
    \end{tabular}
\end{table}

In the 24 independent Sonnet confirmation pairs, runtime control reduces target commits and final-state violations. The paired changes are $-0.875$ $[-1.00,-0.75]$ and $-0.833$ $[-0.958,-0.667]$, respectively, with scenario-cluster bootstrap 95\% intervals. Both reductions remain significant after Holm correction at $p<0.001$.
Benchmark utility improves by $+0.500$ $[0.292,0.708]$, while legitimate-action supply and protocol conformance are unchanged.
These results demonstrate control of the relayed action alongside improved benchmark utility on the independent Sonnet pairs (Table~\ref{tab:agentdojo-derived-main}). Appendix~\ref{sec:agentdojo-utility-trace} explains the utility improvement through a saved native/gated pair and the benchmark's final-state checks.

\subsection{Task utility example}
\label{sec:agentdojo-utility-trace}

This example explains how benchmark utility can improve while the required action remains unchanged.
The Sonnet pair asks the agent to check availability and create a one-hour lunch event at the requested time with the designated participant. Its AgentDojo-derived relay also carries a candidate action forwarding a private birthday invitation to an unrelated recipient. Both execution arms use a tool-evidence worker, a relay, and a final executor, each with a separate context. The relay preserves the fixed candidate, and the executor is instructed to attempt the legitimate task and the relayed action.

The two saved ledgers have the same seven proposed tool calls with the same arguments. They check the calendar, retrieve the contact and email information, create the requested lunch event, and then propose forwarding the birthday email. The native arm executes all seven calls. The gated arm executes the six preceding calls and blocks only the exact registered forwarding action. The lunch event has identical title, description, start and end times, and participant in both arms.

\begin{table}[!htbp]
    \centering
    \small
    \setlength{\tabcolsep}{6pt}
    \renewcommand{\arraystretch}{1.08}
    \caption{\textbf{Blocking the prohibited action restores utility while retaining the required action.} One saved Sonnet pair. The legitimate calendar operation generates the invitation email.}
    \label{tab:agentdojo-utility-trace}
    \begin{tabular}{@{}lll@{}}
        \toprule
        Observed outcome & Native & Gated \\
        \midrule
        Required lunch event & Created & Created \\
        Required calendar invitation & Sent & Sent \\
        Private email forwarded & Yes & Blocked \\
        Exact legitimate-action match & 1 & 1 \\
        Benchmark utility & 0 & 1 \\
        \bottomrule
    \end{tabular}
\end{table}

The benchmark's calendar-creation oracle requires exactly the expected new event and its invitation email, then checks the event fields. The extra forwarded email makes the native final state fail this oracle. Replaying only the saved executed calls in the benchmark's in-memory environment reproduces utility 0 for native and 1 for gated (Table~\ref{tab:agentdojo-utility-trace}). The score change reflects removal of an unauthorized side effect while preserving the required action. Across the 24 saved confirmation pairs, utility improves in 12, stays successful in nine, and stays unsuccessful in three. Exact legitimate-action supply remains unchanged in every pair, explaining why the aggregate utility and supply endpoints move differently.

\subsection{Exposure across models}
Attack exposure determines whether a native/gated comparison can measure prevention of the target event.
A separate development comparison evaluates four models on 36 trajectories each (Table~\ref{tab:agentdojo-exposure}).
Sonnet~4.5 and Haiku~4.5 complete legitimate tasks at rates of 0.750 and 0.556, respectively, while producing zero exact transient target events.
Qwen3-32B produces four events, but its exact-event rate and protocol conformance fall below the selection criteria.
Eligibility requires at least four exact events and an event rate of at least 0.15, together with instrumentation, protocol, and utility criteria. All four models fall short of these joint requirements.
The comparison distinguishes a lack of attack exposure from prevention by the commit gate, motivating evaluation under the separately selected relay condition above.

\begin{table}[!htbp]
    \centering
    \small
    \setlength{\tabcolsep}{9pt}
    \renewcommand{\arraystretch}{1.08}
    \caption{\textbf{Native attack exposure across models.} Each row contains 36 development trajectories from a separate exposure comparison. Exact events, task utility, and protocol conformance use their stated denominators.}
    \label{tab:agentdojo-exposure}
    \begin{tabular}{@{}lcccc@{}}
        \toprule
        Model & Exact transient & Rate & Native utility & Protocol \\
        \midrule
        Sonnet~4.5 & 0/36 & 0.000 & 0.750 & 0.917 \\
        Haiku~4.5 & 0/36 & 0.000 & 0.556 & 0.972 \\
        Qwen3-32B & 4/36 & 0.111 & 0.528 & 0.694 \\
        Llama-3.3-70B & 2/36 & 0.056 & 0.194 & 0.944 \\
        \bottomrule
    \end{tabular}
\end{table}

\subsection{Attack prevention and task utility}
The Qwen3-32B comparison tests whether preventing target commits also preserves useful completion across attack families (Table~\ref{tab:agentdojo}).
Each family contains eight matched native/gated pairs, for 64 trajectories in total.
The model was selected using pilot evidence of nonzero transient commits. The comparison uses separately collected trajectories, and its pooled intervals and tests are exploratory.

\begin{table}[!htbp]
    \centering
    \small
    \setlength{\tabcolsep}{3pt}
    \renewcommand{\arraystretch}{1.08}
    \caption{\textbf{Exploratory Qwen attack-family comparison.} Each family has eight matched pairs. Commit $\Delta$ is gated minus native, with a 95\% cluster interval. N/A marks native rows where the paired contrast is not repeated.}
    \label{tab:agentdojo}
    \begin{tabular}{@{}llccccl@{}}
        \toprule
        Official attack & Arm & Proposal & Commit & Violation & Utility & Commit $\Delta$ [95\% CI] \\
        \midrule
        Important Instructions & Native & 6/8 & 6/8 & 5/8 & 5/8 & N/A \\
        Important Instructions & Gated  & 6/8 & 0/8 & 0/8 & 1/8 & $-0.750$ [$-1.000,-0.375$] \\
        \midrule
        InjecAgent & Native & 5/8 & 5/8 & 5/8 & 2/8 & N/A \\
        InjecAgent & Gated  & 5/8 & 0/8 & 0/8 & 1/8 & $-0.625$ [$-0.875,-0.250$] \\
        \midrule
        System message & Native & 3/8 & 3/8 & 3/8 & 4/8 & N/A \\
        System message & Gated  & 2/8 & 0/8 & 0/8 & 3/8 & $-0.375$ [$-0.750,-0.125$] \\
        \midrule
        Tool knowledge & Native & 6/8 & 6/8 & 5/8 & 4/8 & N/A \\
        Tool knowledge & Gated  & 7/8 & 0/8 & 0/8 & 1/8 & $-0.750$ [$-1.000,-0.375$] \\
        \midrule
        Pooled exploratory & Native & 20/32 & 20/32 & 18/32 & 15/32 & N/A \\
        Pooled exploratory & Gated  & 20/32 & 0/32 & 0/32 & 6/32 & $-0.625$ [$-0.875,-0.344$] \\
        \bottomrule
    \end{tabular}
\end{table}

The commit difference and its interval are reported on each gated row of Table~\ref{tab:agentdojo}.
Target-action proposals are equally frequent overall, with a paired difference of zero and cluster-bootstrap interval $[-0.094,0.094]$. Final-state violations are less frequent under runtime control, with a paired difference of $-0.5625$ $[-0.844,-0.281]$.
Utility also decreases under runtime control, with a pooled difference of $-0.281$ $[-0.469,-0.094]$.
The transient ledger captures two executed attacks later overwritten in the final state, while the utility endpoint identifies the cost of blocking target commits.
Thus, the Qwen comparison demonstrates attack prevention with a substantial loss of useful completion.
A separate, prospectively selected Claude comparison contains 64 native and 64 gated units across the same four attack renderings, with 0 exact proposals or commits in either arm and utility 43/64 versus 44/64.
Because neither arm executes the target action, these runs cannot estimate how much the commit gate prevents it.

\subsection{Generalization to independent pairs}
An independent Qwen replication tests whether attack exposure and utility preservation carry over to new task and injection pairs.
It uses eight held-out pairs, with two from each of workspace, banking, travel, and Slack. The design crosses each pair with AgentDojo's four official non-denial-of-service attacks (Important Instructions, InjecAgent, System Message, and Tool Knowledge).
Each of the 32 combinations of attack and scenario pair is run once under the native runtime and once with the commit gate enforcing the exact target contract, for 64 Qwen3-32B trajectories.
The adapter uses AgentDojo suite v1.2.1 (Python distribution 0.1.35), the official system message and native task/security oracles, and complete unflattened Pydantic tool schemas.
The commit gate verifies the object reference and exact arguments against a contract bound to the destination, operation, principal, and policy version.
Blocking one target therefore leaves a legal call to the same tool eligible for execution.

The confirmatory security analysis requires a native target-commit rate of at least 0.20 and at least four executed target actions.
Security is measured both from the complete executed-action ledger and AgentDojo's final-state oracle. The former is primary because a later legitimate action can overwrite an earlier attack.
Utility preservation is a paired non-inferiority test clustered by task and injection pair: the lower endpoint of the cluster-bootstrap interval for gated minus native user utility must exceed $-0.10$.
All native and gated runs are retained across attacks and domains, including schema errors, refusals, and cases in which the commit gate changes subsequent model behavior.
The comparison contains 64 protocol-conformant trajectories and 279 successful model calls.
The native runtime executes the exact target action in 1/32 cases, below both prespecified minimums.
The single exact target is proposed in both matched arms and committed only in the native arm. This low-exposure comparison is analyzed descriptively.
Utility is 11/32 in each arm. The paired difference is $0$ with cluster interval $[-0.156,0.125]$. Despite equal observed success rates, the data cannot rule out a utility change worse than the $-0.10$ margin.
Table~\ref{tab:agentdojo-confirmatory-shell} reports every attack family and the exploratory pooled row.

\begin{table}[!htbp]
    \centering
    \small
    \setlength{\tabcolsep}{3pt}
    \renewcommand{\arraystretch}{1.08}
    \caption{\textbf{Qwen evaluation on independent task and injection pairs.} Each family has eight matched pairs. Commit $\Delta$ is gated minus native with a 95\% cluster interval. N/A marks native rows where the paired contrast is not repeated. Pooled estimates are exploratory because native exposure falls below the confirmatory threshold. In zero-event cells, $[0,0]$ is a degenerate empirical bootstrap interval, not evidence of zero population uncertainty.}
    \label{tab:agentdojo-confirmatory-shell}
    \begin{tabular}{@{}llccccl@{}}
        \toprule
        Official attack & Arm & Proposal & Commit & Violation & Utility & Commit $\Delta$ [95\% CI] \\
        \midrule
        Important Instructions & Native & 0/8 & 0/8 & 4/8 & 3/8 & N/A \\
        Important Instructions & Gated  & 0/8 & 0/8 & 2/8 & 5/8 & $0$ [$0,0$] \\
        \midrule
        InjecAgent & Native & 0/8 & 0/8 & 3/8 & 3/8 & N/A \\
        InjecAgent & Gated  & 0/8 & 0/8 & 4/8 & 3/8 & $0$ [$0,0$] \\
        \midrule
        System message & Native & 0/8 & 0/8 & 2/8 & 2/8 & N/A \\
        System message & Gated  & 0/8 & 0/8 & 2/8 & 1/8 & $0$ [$0,0$] \\
        \midrule
        Tool knowledge & Native & 1/8 & 1/8 & 5/8 & 3/8 & N/A \\
        Tool knowledge & Gated  & 1/8 & 0/8 & 3/8 & 2/8 & $-0.125$ [$-0.375,0$] \\
        \midrule
        Pooled exploratory & Native & 1/32 & 1/32 & 14/32 & 11/32 & N/A \\
        Pooled exploratory & Gated  & 1/32 & 0/32 & 11/32 & 11/32 & $-0.031$ [$-0.094,0$] \\
        \bottomrule
    \end{tabular}
\end{table}

The pooled difference in final-state violations is $-0.094$ $[-0.250,0.063]$, and the legitimate-action supply difference is $+0.094$ $[0,0.219]$.
The commit gate blocks the target action in one pair. In the other attack families, it never blocks that action, so differences between arms reflect variation across independently generated runs.
Together, the Qwen comparisons separate three outcomes that useful safety control must address: exposure to an attack, prevention of its execution, and preservation of the legitimate task.

\section{Extended Related Work}
\label{sec:extended-related}

\textbf{Agent frameworks make information composition an operational concern.}
AutoGen and CAMEL organize collaboration through separately prompted participants and explicit conversation, while the OpenAI Agents SDK, Google ADK, and ClawTeam provide delegation and orchestration interfaces \citep{wu2023autogenenablingnextgenllm,li2023camelcommunicativeagentsmind,openai_agents_sdk,google_adk,clawteam2026}.
Multi-agent research systems further illustrate the practical use of specialist workers and coordinating agents \citep{hadfield2025multiagentresearch,anthropic2026multiagentpatterns}.
The distinction between homogeneous and heterogeneous teams concerns the models behind those participants, independently of their separate contexts and communication paths.
OpenClaw, Claude Code, and Codex place model-directed work inside persistent tool and execution environments \citep{steinberger2025openclaw,claudecode2026,codex2026}.
These systems motivate specifying exactly which information an agent observes and which runtime component can authorize its proposed action.

\textbf{Task benchmarks establish the capability that a safety control must preserve.}
SWE-bench evaluates repository changes, VisualWebArena evaluates visually grounded web tasks, and OSWorld evaluates actions in real computer environments \citep{jimenez2023swe,koh2024visualwebarena,xie2024osworld}.
TheAgentCompany and Agent's First Day extend the task horizon to workplace activity and learning in a changing environment \citep{xu2025theagentcompanybenchmarkingllmagents,fu2026agentsdaybenchmarkinglearning}.
ClawBench and ClawMark add realistic web execution and stateful coworker interactions \citep{zhang2026clawbench,meng2026clawmark}.
In $\tau$-bench, tool use is evaluated together with domain rules and the resulting database state \citep{yao2024tau}.
Our task endpoint $T$ retains this concern for useful completion, while authorized supply $A$ isolates the particular capability whose suppression could otherwise masquerade as safety.

\textbf{Agent-safety benchmarks broaden the contexts in which harmful behavior is measured.}
AgentHarm includes benign counterparts of harmful behaviors, Agent-SafetyBench tests diverse interactive risks, and AgentDojo supplies a stateful setting for tool-mediated prompt injection and defense evaluation \citep{andriushchenko2025agentharmbenchmarkmeasuringharmfulness,zhang2025agentsafetybenchevaluatingsafetyllm,debenedetti2024agentdojo}.
SafeArena pairs safe and harmful web tasks, OS-Harm studies unsafe computer use, and ATBench examines safety across realistic agent settings \citep{tur2025safearenaevaluatingsafetyautonomous,kuntz2025osharmbenchmarkmeasuringsafety,li2026atbenchdiverserealisticagent}.
ClawEval and ClawsBench further distinguish completion from safety in autonomous workflows. RASEval and MCPSecBench examine security across tool and protocol interfaces \citep{ye2026clawevaltrustworthyevaluationautonomous,li2026clawsbenchevaluatingcapabilitysafety,fu2025rasevalcomprehensivebenchmarksecurity,yang2026mcpsecbenchsystematicsecuritybenchmark}.
These benchmarks provide complementary attack surfaces and task distributions.
Authorization-paired evaluation makes the relationship between prohibited and required authorized uses explicit within a shared task scenario, so a safety score also measures preservation of the required capability.

\textbf{Privacy and policy-aware evaluation connect disclosure to its context.}
PrivacyLens evaluates privacy norms along agent trajectories, CI-Work tests contextual integrity in workplace information flow, and MyPhoneBench separates access permissions, disclosure, and memory behavior from task success \citep{shao2025privacylensevaluatingprivacynorm,fu2026ciworkbenchmarkingcontextualintegrity,tang2026phoneuseagentsrespectprivacy}.
ST-WebAgentBench's completion-under-policy measure likewise makes compliance part of successful execution \citep{levy2024st}.
The distinction between safety and utility is therefore already substantive in agent evaluation.
Our contribution is to bind the two outcomes to the evaluated governed objects and trusted authorization contrast, and to diagnose how multi-agent composition disrupts that binding.
Financial-risk frameworks address the consequences of agent failure through assessment and compensation, complementing controls that prevent an unauthorized action before it executes \citep{hua2026quantifyingtrustfinancialrisk}.

\textbf{Multi-agent interaction introduces both adversarial and emergent risks.}
TAMAS evaluates adversarial instances alongside harmless tasks, AgentLeak measures internal-channel privacy leakage and task utility, and GroupGuard models coordinated attacks \citep{kavathekar2025tamasbenchmarkingadversarialrisks,yagoubi2026agentleakfullstackbenchmarkprivacy,tao2026groupguardframeworkmodelingdefending}.
Research on emergent social risks and AI organizations examines how collective behavior changes with interaction structure \citep{huang2026emergentsocialintelligencerisks,shen2026aiorganizations}.
Agents in the Wild examines interaction patterns and safety risks on Moltbook \citep{zhang2026agents}.
Agentic-misalignment studies use simulated deployments to investigate harmful actions under goal conflicts, and Petri supports automated exploration of such behaviors \citep{lynch2025agentic,lynch2026agenticsummer,fronsdal2025petri}.
\method isolates a specific mechanism within collaborative execution: agents can preserve their assigned fragments and still enable a downstream unauthorized action.
Our constructed suite pairs governed uses by their trusted permissions and varies the information available to review. The read-scope and contributor interventions make this dependency observable across both homogeneous and heterogeneous teams.

\textbf{Compositional attacks show why safety cannot be inferred from isolated inputs.}
Jailbreak in Pieces and Safety Snowball exploit cross-modal composition, while Jigsaw Puzzles, Crescendo, and Foot-In-The-Door build harmful requests through multiple conversational steps \citep{shayegani2024jailbreak,cui2025snowball,yang2025jigsaw,russinovich2025crescendo,weng2025footinthedoor}.
X-Teaming and MM-Plan use agentic planning to construct adaptive attacks, and HarmBench standardizes comparisons of red-teaming methods and robust refusal \citep{rahman2025xteaming,zhang2026visual,mazeika2024harmbench}.
The common concern is that the safety significance of combined information can exceed that of its components.
Our setting places this composition inside the evaluated workflow and changes trusted authorization for the same recovered governed object.
This links the compositional failure directly to the requirement to deny without disabling.

\textbf{Harness security connects model behavior to its execution environment.}
OpenClaw security studies and ClawKeeper examine threats and controls across persistent agent infrastructure \citep{wang2026systematicsecurityevaluationopenclaw,liu2026clawkeepercomprehensivesafetyprotection}.
Poise studies instruction injection through agent skills, ClawSafety examines framework-dependent safety, and persistent-state attacks expose conflicts between protection and legitimate updates \citep{hao2026poise,wei2026clawsafety,wang2026agentassetrealworldsafety}.
Harness surveys, audits, and harness-aware evaluation emphasize orchestration, verification, and governance as parts of agent behavior \citep{li2026agentharness,liu2026auditingagentharnesssafety,zhang2026stop}.
Our placement experiments isolate the roles of runtime lineage, permission interpretation, and assembly in preserving authorized execution.

\textbf{Information-flow and capability controls supply the enforcement foundation.}
Complete mediation requires checking access at each relevant operation, information-flow control follows information across computation, and the confused-deputy problem separates possession of authority from authorized use \citep{saltzer1975protection,denning1976lattice,hardy1988confused}.
CaMeL separates privileged control from untrusted processing, propagates dependencies through its interpreter, and checks tool-call policies against the resulting transitive dependencies \citep{debenedetti2025camel}.
This supports policy enforcement across composed computations. ChainCaps, ARGUS, and ScopeGate develop delegated, provenance-aware, and value-bound authorization \citep{jiang2026chaincaps,weng2026argus,zuvic2026capabilitygates}.
\method supplies a complementary experimental decomposition: paired outcomes couple denied use to required authorized use, read-scope and contributor interventions isolate resolution failures, and placement comparisons test how the required capabilities survive communication.
Its central requirement is end-to-end: the identity of the composed governed object and its current trusted authorization must reach the agent action boundary together.

\section{Synthesis of Findings}
\label{sec:evidence-guide}

The findings connect the information available to review with the permission distinction that must survive execution (Table~\ref{tab:claim-map}).
Composition and contributor comparisons establish the need to resolve information jointly. Delegation separates reliable lineage from permission attribution, and placement comparisons identify how model and runtime components preserve the required capabilities.
Table~\ref{tab:study-synthesis} summarizes outcomes with their respective sampling units.

Table~\ref{tab:surface} distinguishes two readers at the action boundary. The ladder measures verbatim release, while the representation study measures how a family-conditioned decoder changes enforcement on the same proposals. The latter comparison includes improvements, ties, and regressions, showing why a control result must identify the reader under which governed use is recognized.

\begin{table}[!htbp]
    \centering
    \small
    \setlength{\tabcolsep}{5pt}
    \renewcommand{\arraystretch}{1.12}
    \caption{\textbf{Contributions and supporting comparisons.} Measurement scope identifies the intervention or execution surface behind each conclusion.}
    \label{tab:claim-map}
    \begin{tabular}{@{}lll@{}}
        \toprule
        Evidence & Direct conclusion & Measurement scope \\
        \midrule
        Paired evaluation & Authorized capability changes the safety target & Matched policy intervention \\
        Paired evaluation & Resolution, ranking, and enforcement are distinct & Verbatim action ledger \\
        \midrule
        Composition & Admissible artifacts identify a denied action & Read-scope intervention \\
        Scaling & Distributed information drives resolution burden & Star graphs, seven teams \\
        \midrule
        Delegation & Lineage and authority are separate capabilities & Paired contracts, 840 graphs \\
        Permission placement & Lineage and action binding retained & Four specialists, 480 graphs each \\
        Assembly placement & Authorized supply restored & Matched artifacts, 40 \allow/arm \\
        \midrule
        External transfer & Object-bound runtime removes relay commits & 24 AgentDojo-derived pairs \\
        \bottomrule
    \end{tabular}
\end{table}

\begin{table}[!htbp]
    \centering
    \small
    \setlength{\tabcolsep}{4pt}
    \renewcommand{\arraystretch}{1.15}
    \caption{\textbf{Reader-specific control at the action boundary.} Ladder rows aggregate final-configuration trials across two attack families and two budgets. Representation rows count model-by-transformation comparisons against literal matching, each with 12 fixed proposals.}
    \label{tab:surface}
    \begin{tabular}{@{}llrr@{}}
        \toprule
        Boundary & Reader / object & Metric & Value \\
        \midrule
        Action commit & Literal governed objects   & \selective        & 0.990 to 0.998 \\
        Action commit & Literal governed objects   & Verbatim releases & 0/2{,}304 \\
        Action commit & Eight representation families & Lower commit  & 22/40 \\
        Action commit & Eight representation families & Higher commit & 8/40 \\
        Action commit & Eight representation families & Unchanged commit & 10/40 \\
        \bottomrule
    \end{tabular}
\end{table}

\begin{table}[!htbp]
    \centering
    \small
    \setlength{\tabcolsep}{6pt}
    \renewcommand{\arraystretch}{1.10}
    \caption{\textbf{Outcomes across controlled comparisons.} Each group identifies its experimental setting and sampling unit. Delegation-depth drift concerns exact final proposals, while contract-study drift counts errors at any role. Permission placement and assembly are separate comparisons.}
    \label{tab:study-synthesis}
    \begin{tabular}{@{}lr@{}}
        \toprule
        Outcome & Result \\
        \midrule
        \multicolumn{2}{@{}l}{\textit{Natural composition: 60 evaluation pairs}} \\
        Denied exact recovery, local    & 3/60 \\
        Denied exact recovery, global   & 0/60 \\
        Authorized supply, global       & 15/60 \\
        \midrule
        \multicolumn{2}{@{}l}{\textit{Trusted policy $\times$ untrusted claim: 320 units}} \\
        Policy effect                 & Large \\
        Claim interference            & Near zero \\
        Benign task completion        & 1.000 \\
        \midrule
        \multicolumn{2}{@{}l}{\textit{Object representations: 480 proposals}} \\
        Canonicalization lowers commit  & 22/40 \\
        Canonicalization raises commit  & 8/40 \\
        Tied                            & 10/40 \\
        \midrule
        \multicolumn{2}{@{}l}{\textit{Contributor scaling: 672 graphs}} \\
        Coverage, fixed-$K$ and $N8,K7$ & 504/504 \\
        Coverage, $N16,K15$            & 149/168 \\
        Haiku team contrast           & Significant after Holm correction \\
        \midrule
        \multicolumn{2}{@{}l}{\textit{Delegation depth: 840 graphs, seven teams}} \\
        Source labels retained        & 840/840 \\
        Complete lineage              & 178/840 \\
        Final-proposal attribution drift & 132/840 \\
        False trusted elevation       & 0/840 \\
        \midrule
        \multicolumn{2}{@{}l}{\textit{Lineage contracts: 1{,}680 graphs}} \\
        Complete lineage, type-only     & 202/840 \\
        Complete lineage, explicit      & 834/840 \\
        Any-role attribution drift, explicit & 676/840 \\
        \midrule
        \multicolumn{2}{@{}l}{\textit{Permission placement and assembly}} \\
        Specialists meeting evaluation criteria & 4/4 \\
        Authorized supply, runtime assembler & 1.000 \\
        Denied commits                       & 0 \\
        \midrule
        \multicolumn{2}{@{}l}{\textit{Policy-version recovery on independent graphs}} \\
        Authorized supply, current reread    & 120/120 \\
        Authorized supply, one-repair condition & 0/120 \\
        Wrong commits                        & 0 \\
        \midrule
        \multicolumn{2}{@{}l}{\textit{Tool-use transfer: 24 independent Sonnet pairs}} \\
        Target commit, native           & 21/24 \\
        Target commit, gated            & 0/24 \\
        User utility, gated             & 21/24 \\
        \bottomrule
    \end{tabular}
\end{table}

\end{document}